\documentclass[fleqn,usenatbib]{mnras}

\usepackage{newtxtext,newtxmath}

\usepackage[T1]{fontenc}
\usepackage{ae,aecompl}

\usepackage{graphicx}	
\usepackage{amsmath}	
\usepackage{svg}
\usepackage{gensymb}
\usepackage{xspace}
\usepackage[export]{adjustbox} 
\usepackage{xcolor}
\usepackage{orcidlink}
\usepackage{soul}

\providecommand{\bjdtdb}{\ensuremath{\rm {BJD_{TDB}}}}

\providecommand{\lst}{\ensuremath{\,L_\odot}}
\providecommand{\mj}{\ensuremath{\,{\rm M_J}}}
\providecommand{\rj}{\ensuremath{\,{\rm R_J}}}

\providecommand{\mst}{\ensuremath{\,{\rm M_\odot}}}
\providecommand{\rst}{\ensuremath{\,{\rm R_\odot}}}
\providecommand{\fave}{\langle F \rangle}
\providecommand{\fluxcgs}{10$^9$ erg s$^{-1}$ cm$^{-2}$}
\providecommand{\arcsec}{$^{\prime \prime}$}
\providecommand{\tess}{{\it TESS}}
\newcommand{\angstrom}{\mbox{\normalfont\AA}}

\graphicspath{{./}{figures/}}

\newcommand\dsout[1]{}

\title[A transiting brown dwarf--M-dwarf system]{TOI-2119: A transiting brown dwarf orbiting an active M-dwarf from NASA's {\it TESS} mission}

\author[Carmichael et al.]{Theron W. Carmichael, $^1$\thanks{E-mail: tcarmich@ed.ac.uk}\orcidlink{0000-0001-6416-1274}
Jonathan M. Irwin, $^2$
Felipe Murgas, $^{3,4}$
Enric Pall\'{e}, $^{3,4}$\orcidlink{0000-0003-0987-1593}\newauthor 
Keivan G.\ Stassun, $^{5,6}$\orcidlink{0000-0002-3481-9052}
Matthew Bartnik, $^{20}$ 
Karen A.\ Collins, $^2$\orcidlink{0000-0001-6588-9574} 
Jerome de Leon, $^8$\orcidlink{0000-0002-6424-3410}\newauthor 
Emma Esparza-Borges, $^{3,4}$\orcidlink{0000-0002-2341-3233} 
Jeremy Fedewa, $^{20}$ 
William Fong, $^{15}$
Akihiko Fukui, $^{9,3}$\orcidlink{0000-0002-4909-5763}\newauthor 
Jon M. Jenkins, $^{18}$ \orcidlink{0000-0002-4715-9460}
Taiki Kagetani, $^{10}$\orcidlink{0000-0002-5331-6637}
David W. Latham, $^2$
Michael B. Lund,$^{7}$\orcidlink{0000-0003-2527-1598}\newauthor
Andrew W. Mann, $^{11}$\orcidlink{0000-0003-3654-1602}
Dan Moldovan, $^{17}$ 
Edward H. Morgan, $^{15}$ 
Norio Narita, $^{9,3,12}$\orcidlink{0000-0001-8511-2981}\newauthor 
Shane Painter,$^{20}$ 
Hannu Parviainen, $^{3,4}$\orcidlink{0000-0001-5519-1391} 
Elisa V. Quintana, $^{16}$\orcidlink{0000-0003-1309-2904}
George R. Ricker, $^{15}$\orcidlink{0000-0003-2058-6662} \newauthor 
Jack Schulte, $^{20}$ 
Richard P. Schwarz, $^{13}$ \orcidlink{0000-0001-8227-1020}
Sara Seager, $^{15}$ 
Kirill Sokolovsky, $^{20}$\orcidlink{0000-0001-5991-6863}\newauthor 
Joseph D. Twicken, $^{18,19}$ \orcidlink{0000-0002-6778-7552}
Joshua N. Winn $^{14}$ \orcidlink{0000-0002-4265-047X}
\\
$^1$ Institute for Astronomy, University of Edinburgh, Royal Observatory, Blackford Hill, Edinburgh, EH9 3HJ, UK\\
$^2$ Center for Astrophysics, Harvard \& Smithsonian, 60 Garden Street, Cambridge, MA 02138, USA\\
$^3$ Instituto de Astrof\'{i}sica de Canarias, 38200 La Laguna, Tenerife, Spain\\
$^4$ Department Astrof\'{i}sica, Universidad de La Laguna, 38206 La Laguna, Tenerife, Spain\\
$^5$ Vanderbilt University, Department of Physics \& Astronomy, 6301 Stevenson Center Ln., Nashville, TN 37235, USA\\
$^6$ Fisk University, Department of Physics, 1000 18th Ave. N., Nashville, TN 37208, USA\\
$^7$ Caltech IPAC--NASA Exoplanet Science Institute 1200 E. California Ave, Pasadena, CA 91125, USA\\
$^8$ Department of Astronomy, Graduate School of Science, The University of Tokyo, 7-3-1 Hongo, Bunkyo-ku, Tokyo 113-0033, Japan\\ 
$^9$ Komaba Institute for Science, The University of Tokyo, 3-8-1 Komaba, Meguro, Tokyo 153-8902, Japan\\
$^{10}$ Department of Multi-Disciplinary Sciences, Graduate School of Arts and Sciences, The University of Tokyo, 3-8-1 Komaba, \\ ~~~~Meguro, Tokyo 153-8902, Japan\\
$^{11}$ Department of Physics and Astronomy, The University of North Carolina at Chapel Hill, Chapel Hill, NC 27599-3255, USA\\
$^{12}$ Astrobiology Center, 2-21-1 Osawa, Mitaka, Tokyo 181-8588, Japan\\
$^{13}$ Patashnick Voorheesville Observatory, Voorheesville, NY 12186, USA\\
$^{14}$ Department of Astrophysical Sciences, Princeton University, 4 Ivy Lane, Princeton, NJ 08544, USA\\
$^{15}$ Department of Physics, and Kavli Institute for Astrophysics and Space Research, Massachusetts Institute of Technology, Cambridge, MA 02139, USA\\
$^{16}$ NASA Goddard Space Flight Center, 8800 Greenbelt Road, Greenbelt, MD 20771, USA\\
$^{17}$ Google, Cambridge, MA, USA\\
$^{18}$ NASA Ames Research Center, Moffett Field, CA 94035, USA\\
$^{19}$ SETI Institute, Mountain View, CA 94043, USA\\
$^{20}$ Department of Physics and Astronomy, Michigan State University, East Lansing, MI 48824, USA\\
}

\date{Accepted XXX. Received YYY; in original form ZZZ}

\pubyear{2022}

\begin{document}
\label{firstpage}
\pagerange{\pageref{firstpage}--\pageref{lastpage}}
\maketitle

\begin{abstract}
    \noindent We report the discovery of TOI-2119b, a transiting brown dwarf (BD) that orbits and is completely eclipsed by an active M-dwarf star. Using light curve data from the Transiting Exoplanet Survey Satellite mission and follow-up high-resolution Doppler spectroscopic observations, we find the BD has a radius of $R_b = 1.08 \pm 0.03{\rm R_J}$, a mass of $M_b = 64.4 \pm 2.3{\rm M_J}$, an orbital period of $P = 7.200865 \pm 0.00002$ days, and an eccentricity of $e=0.337\pm 0.002$. The host star has a mass of $M_\star = 0.53 \pm 0.02{\rm M_\odot}$, a radius of $R_\star= 0.50 \pm 0.01{\rm R_\odot}$, an effective temperature of $T_{\rm eff} = 3621 \pm 48$K, and a metallicity of $\rm [Fe/H]=+0.06\pm 0.08$. TOI-2119b joins an emerging population of transiting BDs around M-dwarf host stars, with TOI-2119 being the ninth such system. These M-dwarf--brown dwarf systems typically occupy mass ratios near $q = M_b/M_\star \approx 0.1-0.2$, which separates them from the typical mass ratios for systems with transiting substellar objects and giant exoplanets that orbit more massive stars. The nature of the secondary eclipse of the BD by the star enables us to estimate the effective temperature of the substellar object to be $2030\pm 84$K, which is consistent with predictions by substellar evolutionary models.
\end{abstract}



\section{Introduction} \label{sec:intro}
Astronomers traditionally define BDs as the objects between giant planets and stars, occupying the mass range $13-80$ Jupiter masses ($\mj$) based on deuterium fusion (at $13\mj$) and hydrogen fusion (at $80\mj$) thresholds. \textcolor{black}{However, metallicity and other factors may change these boundaries in the range $11-16\mj$ \citep{spiegel2011} and $75-80\mj$ \cite{baraffe2002}, respectively.} From this, a simple mass-based definition does not tell us whether BDs can form like giant planets, low-mass stars, or a mixture of both populations' formation pathways. We are now poised to more thoroughly explore this question of formation mechanism by examining the emerging transiting BD population with the Transiting Exoplanet Survey Satellite (\tess) mission \citep{tess}. Transiting BDs are key as they enable us to measure the BD's mass and radius, which typically ranges from $0.7-1.4$ Jupiter radii ($\rj$) around main sequence host stars \citep{corot3b, subjak2019, carmichael2020}. Though transit detections typically restrict us to exploring a parameter space for short-period BDs (usually $P_{\rm orb} < 30$ days to efficiently detect and characterise multiple transits), they play a critical role in filling out the population of BDs with measured masses and surface gravity ($\log{g}$) estimates. The mass and radius are key parameters involved in the construction of BD evolutionary models \citep{chabrier97,baraffe03, saumon08, ATMO2020, sonora21}, to which we compare the observed population of transiting BDs. This improves our understanding of the evolution of these objects and with a large enough sample size of BDs with known masses, radii, and ages (from age dating techniques of the host stars), we stand to learn more about how the dominant formation mechanism changes for objects from giant planets to low-mass stars as a function of mass.

The \tess\, mission has been an invaluable resource in this endeavor. After delivering several new transiting BD discoveries during its two-year primary mission \citep[e.g.][]{jackman2019, subjak2019, benni2020, carmichael2020, carmichael2021}, \tess\, continues to serve as our primary detection tool for new transiting BDs in its extended missions \citep[e.g.][]{artigau2021, grieves2021}. The key parameters from the \tess\, light curve data are the radius, orbital period, and orbital inclination of transiting BDs. The radius, which cannot be known to a precision greater than that of the host star, establishes which transit events are candidate transiting BDs. As mentioned, the range of radii we consider for such BD candidates typically falls between $0.7-1.4{\rm R_J}$ unless the candidate orbits a pre-main sequence host star or a star that is otherwise known to be particularly young ($<100$ Myr). \textcolor{black}{In such cases, larger BD candidate radii up to $3-5\rj$ are considered. These larger radii are considered because the largest known transiting BD is RIK 72b with a radius of $3.1\rj$ around a 5 Myr old star in Upper Scorpius \citep{david19_bd}. Once an estimate for the radius of a BD candidate is obtained to fall in this range, we then pursue radial velocity (RV) follow up for a mass determination.}

The orbital period establishes the appropriate timing for follow up RV observations, which come in the form of spectra of the host star, and are most useful for determining an initial estimate for the mass of the candidate transiting BD as possible. As we use these spectra to construct an orbital solution for the star, we also derive the spectroscopic effective temperature, metallicity ([Fe/H]), and rotational broadening of the star (specifically, $v\sin{I_\star}$). The orbital solution alone gives an estimate of the minimum mass ($m\sin{i}$), which is not always close to the true mass $M_b$ of the companion, especially for systems observed at a low orbital inclination. By making use of the orbital inclination $i$ obtained from the transit data, we break the degeneracy between the companion mass and orbital inclination.

The precision of these mass and radius measurements is also an important consideration. \textcolor{black}{The RV amplitude signal for a short-period BD orbiting a typical FGK main sequence star is on the order of several kilometers per second. This is based on the constraints that BDs are defined to be $13-80\mj$, FGK stars are roughly $0.6-1.5\mst$, and the orbital periods are $\lesssim 30$ days; see \cite{fischer2014_techniques} for the equation used to quickly estimate this}. Modern echelle spectrographs with resolving power on the order of $R\approx 10^4$ are well within their capabilities for precisely measuring RVs in this range. This is especially true given the selection of relatively bright ($V<12$) stars that the \tess\, mission observes. The precision of radius measurements is limited by several factors: the signal-to-noise per transit, the number of transits observed, the precision of the orbital impact parameter, and the precision of stellar radius measurement. The signal-to-noise per transit criterion is met by the sensitivity of the \tess\, mission when it observes bright enough stars. 

The second limiting factor is the orbital impact parameter $b$, which is degenerate with the radius of the BD when the signal-to-noise is low or when $b$ approaches a value of 1. We do not have control of this as it is based on the orbital geometry of the transiting BD relative to its host star. The third bottleneck is the precision to which we determine the radius of the host star. This is addressed with the use of the parallax measurements released in the Gaia mission's early data release 3 (Gaia EDR3, \cite{edr3}). These parallax measurements typically play the most important role in reducing the radius uncertainty for stars. These measurements are directly related to stellar distances, which we use to determine stellar luminosities that are used to better constrain the evolutionary state of the star.

This brings us to the most important parameter when characterising the host star: stellar age. We assume any star hosting a transiting BD shares its age with the BD, meaning that the system's age is valuable in testing substellar evolutionary models for the BD. One way to do this testing is in mass--radius space for BDs and principally involves testing how well theoretical and simulation-driven works \citep[e.g.][]{baraffe03, saumon08, ATMO2020, sonora21} predict the rate at which the radius of BDs decreases with age. We generally expect BDs to contract with age as they cool over time as they lack a mechanism to initiate any increase in radius like stars do. The rate of this contraction is predicted to be greater at younger ages and thought to decelerate at BD ages beyond a few billion years, so it is essential that we determine a precise age to conduct meaningful tests of this prediction. Though a precise determination of both the BD's radius and the star's age are crucial in this regard, we more often find ourselves challenged to reliably estimate the age of the host star. 

Recently, M-dwarf stars have begun to emerge as common hosts to transiting BDs \citep{david19_bd, jackman2019, palle2021, acton2021}. That is to say, of the roughly 30 known transiting BDs, M-dwarfs are hosts to 9 of them. \textcolor{black}{The numbers for A, F, G, and K-dwarf host stars are 2, 9, 9, and 1, respectively}. M-dwarf stars are relatively difficult to characterise compared to their FGK main sequence counterparts, especially when estimating the stellar age. This is because M-dwarfs change very little once they have entered the main sequence and they remain this way for much longer than their typical FGK stellar counterparts. However, recent works by \cite{mann2015} and \cite{engle2018} have made this art of charactersing some of the fundamental properties of M-dwarfs (mass, radius, effective temperature, and even age) more reliable. The empirical M-dwarf relationships explored in works like these facilitate a more reliable determination of M-dwarf properties such that we may use them for a better understanding of the properties of transiting BDs that we may find orbiting them.

\begin{figure*}
\centering
\includegraphics[width=0.95\textwidth, trim={0.0cm 0.0cm 0.0cm 0.0cm}]{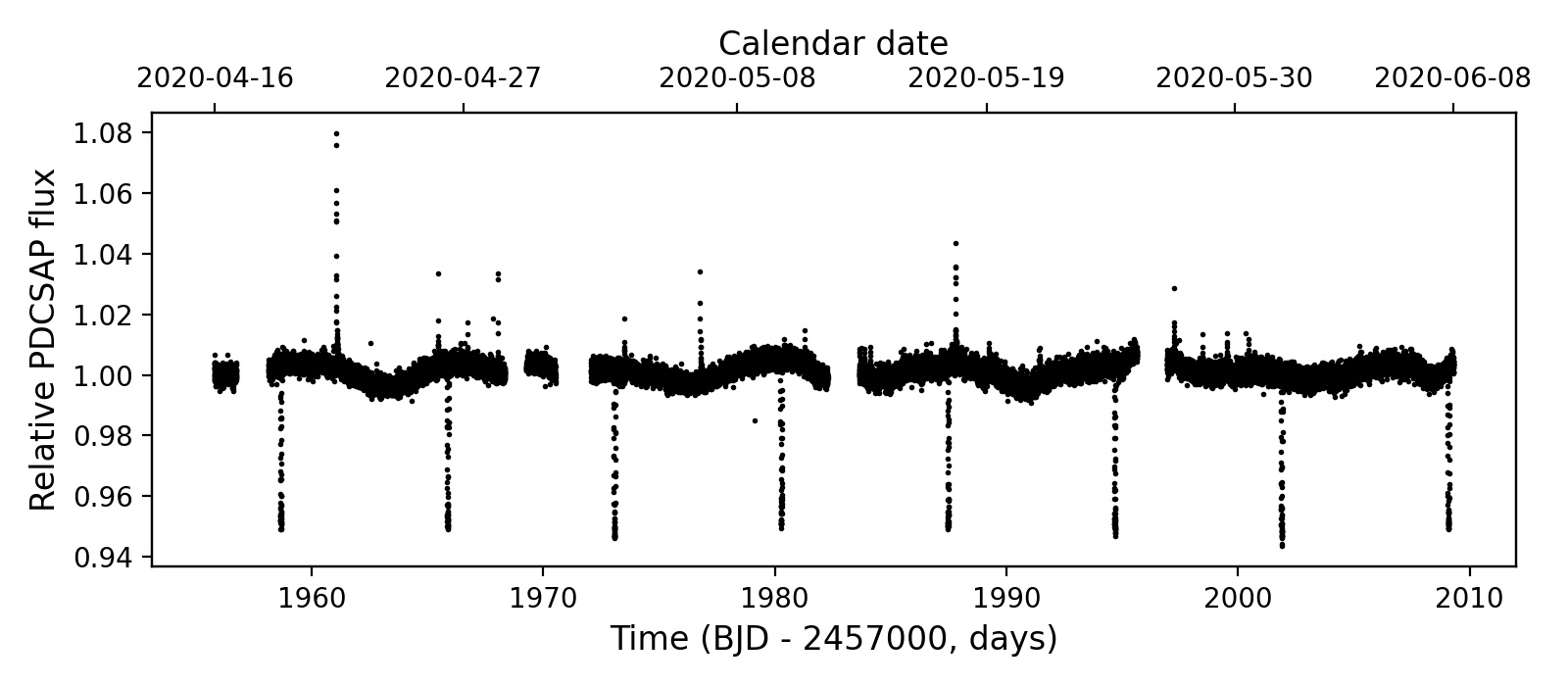}    
\caption{PDCSAP \tess\, light curve of TOI-2119. This star appears to be relatively active given the number and strength of the flares observed in the \tess\, data over 2 sectors.}\label{fig:tess_lc}
\end{figure*}

In this work, we present results, analysis, and discussion on the TOI-2119 (TIC 236387002) system. This system contains a transiting BD, TOI-2119b, in a short-period orbit ($P = 7.2$ days) around an early-type M-dwarf star. This transiting BD is completely eclipsed by the M-dwarf. Section \ref{sec:observations} presents the \tess\, light curves and other follow up data used to derive the physical properties of the star and BD. Section \ref{sec:analysis} discusses the analysis tools used to derive these parameters and any caveats given the difficult nature of determining the mass, radius, and age of low-mass stars. Section \ref{sec:conclusion} focuses on the substellar mass--radius diagram and the usefulness of TOI-2119b in testing substellar evolutionary models.


\section{Observations}\label{sec:observations}
\subsection{{\it TESS} light curves}\label{sec:tess_lc}
The \tess\, mission produced the light curves for TOI-2119 (TIC 236387002) with a two-minute cadence in sector 24 from 16 April 2020 to 12 May 2020 and in sector 25 from 14 May 2020 to June 2020. A transit signature with a 7.2-day period was identified in the Science Processing Operations Centre \citep[SPOC;][]{jenkins_2016} transiting planet search \citep{jenkins2002, jenkins2010, jenkins2020} of the light curves in both sectors. This SPOC detection was subsequently promoted by the \tess\, Science Office to \tess\, Object of Interest \citep[TOI;][]{guerrero2021} status as TOI 2119.01 based on the clean transit model fit and diagnostic test results in the SPOC data validation report \citep{Twicken:DVdiagnostics2018, Li:DVmodelFit2019}.

These data are from the Presearch Data Conditioning Simple Aperture Photometry flux \citep[PDCSAP;][]{stumpe2014_pdc, smith2012_pdc} available through the Mikulski Archive for Space Telescopes (MAST). The PDCSAP light curve removes some systematic stellar effects but aims to retain native stellar features such as brightness modulation, transits, eclipses, and flares (see Figure \ref{fig:tess_lc}), \textcolor{black}{so we use the PDCSAP data for the transit analysis.} We then use the {\tt lightkurve} \citep{lightkurve} package in Python to normalise the light curve. We manually remove the flares from the light curve data and note here that roughly two dozen flares brighten more than $0.5\%$ above baseline. At this step, the light curve data are ready for transit and secondary eclipse analysis.

\subsubsection{Photometric modulation from the TESS data}
\begin{figure}
\centering
\includegraphics[width=0.45\textwidth, trim={0.0cm 0.0cm 0.0cm 0.0cm}]{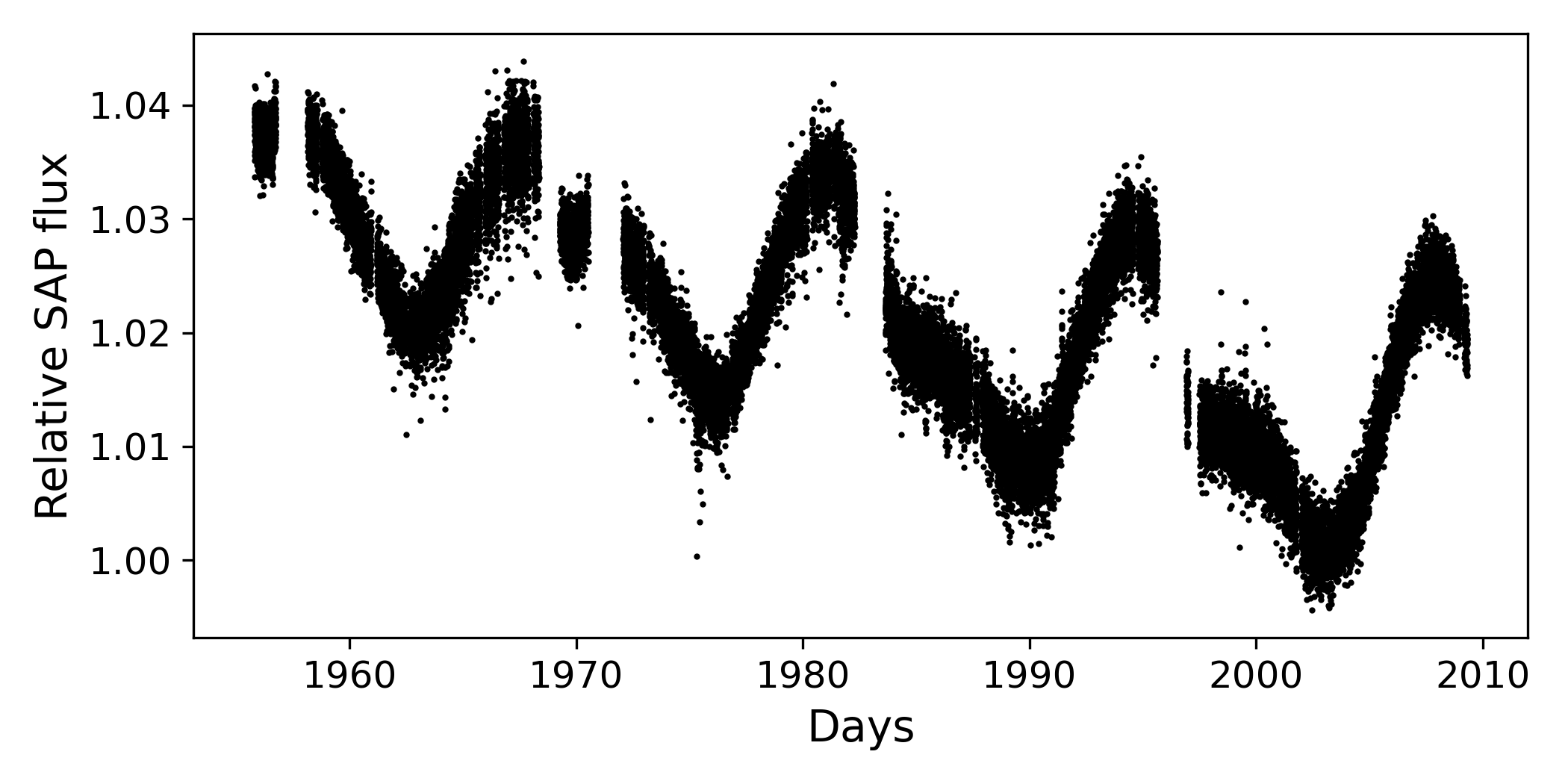}    
\includegraphics[width=0.45\textwidth, trim={0.0cm 0.0cm 0.5cm 0.0cm}]{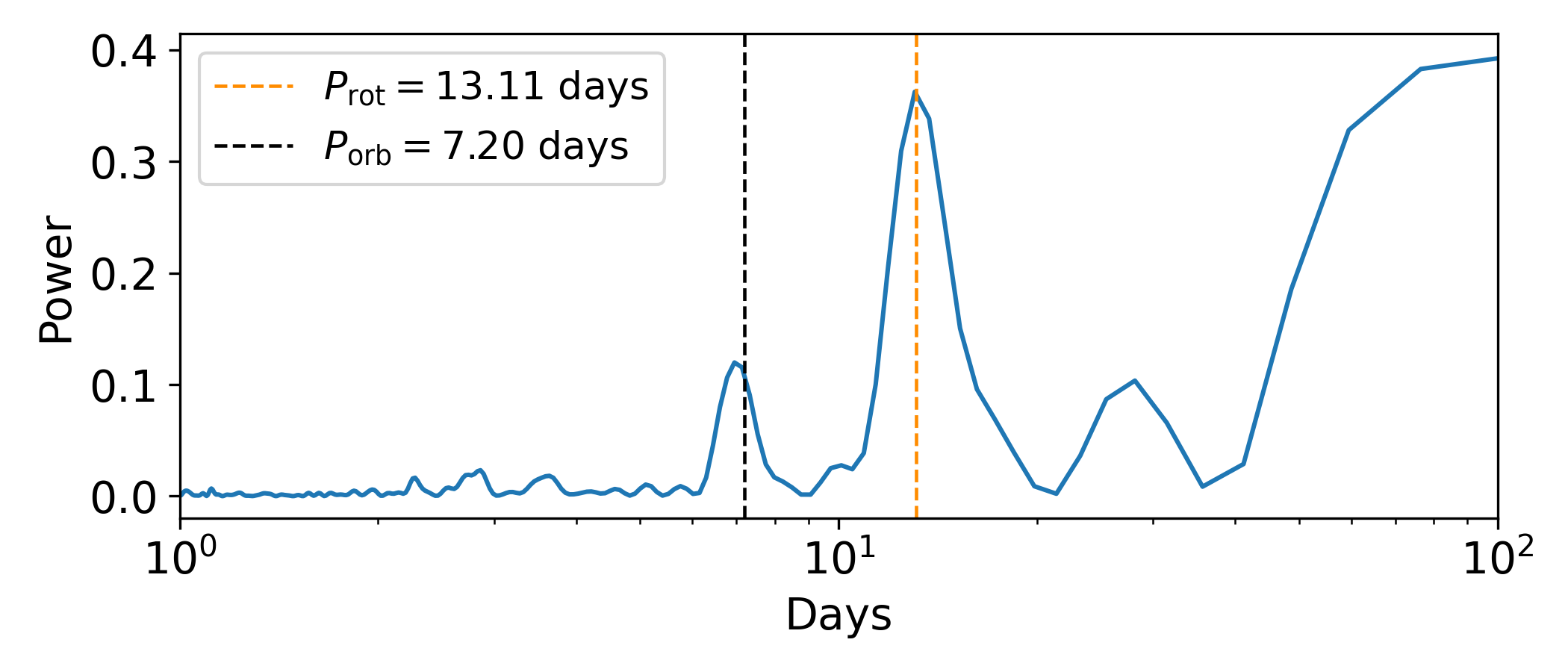}
\includegraphics[width=0.45\textwidth, trim={0.75cm 0.0cm 0.5cm 0.0cm}]{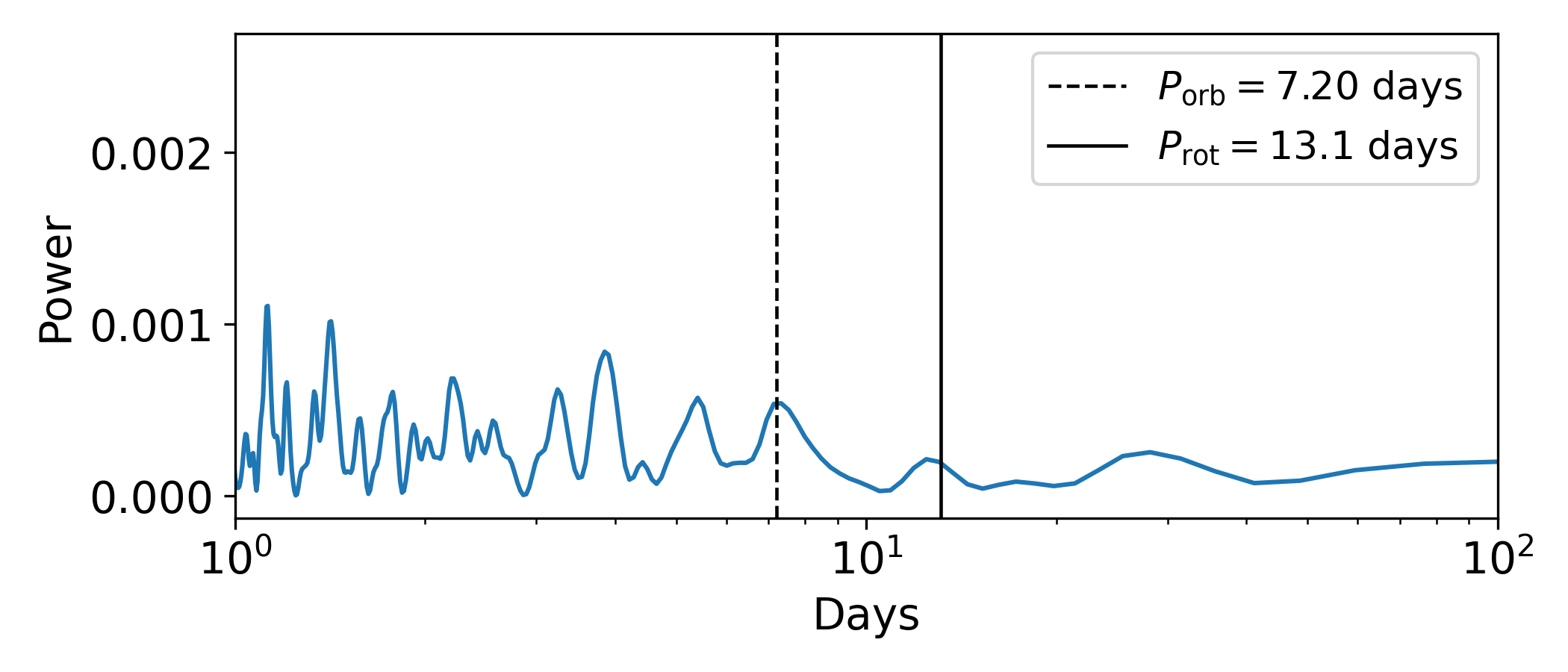}
\caption{Top: \tess\, SAP light curve of TOI-2119 with flares and transits removed. Middle: Lomb-Scargle periodogram of the data shown in the top panel. The value of and uncertainty on the rotation period ($P_{\rm rot} = 13.11\pm 1.41$ days at 1$\sigma$) are estimated by fitting a simple Gaussian function to the peak in the periodogram (ignoring the the power at $10^2$ days due to the long-term downward trend in the light curve data). The peaks at approximately 6.9 days ($P_{\rm rot}/2$) and 26.9 days ($P_{\rm rot}\times 2$) appear to be harmonics of the stellar rotation period and are not caused by the transit events (as they have been removed). \textcolor{black}{Bottom: Periodogram analysis of normalised \tess\ data with transits and eclipses removed, but with the stellar flares retained. There is no evidence of strong periodicity or quasi-periodicity from the flares alone.}}\label{fig:periodogram}
\end{figure}

The SPOC Simple Aperture Photometry \citep[SAP;][]{twicken:PA2010SPIE, morris2020_sap} flux data reveal photometric modulation on the order of a few percent from TOI-2119. \textcolor{black}{We use the SAP data over the PDCSAP when examining photometric modulation as the PDC algorithm is sometimes prone to suppressing the amplitude of the variability in the light curve data.} To characterise the periodicity of this modulation, we first mask the transits of TOI-2119b and the flares before applying a Lomb-Scargle periodogram analysis of this data to estimate a 13.11-day period (Figure \ref{fig:periodogram}). We attribute this to a stellar rotation rate of $P_{\rm rot} = 13.11 \pm 1.41$ days caused by star spots moving in and out of our line of sight on the stellar surface. The uncertainty on the rotation period is estimated from the width of a simple Gaussian function that we fit to the peak in the periodogram. There also appears to be a slight downward trend in the star's brightness over a much longer period than we find here, but the \tess\, data do not span a sufficient baseline to explore this further.

\subsubsection{Flares in the M-dwarf light curve}
\textcolor{black}{We follow \cite{medina2020} in our definition of a flare event as three consecutive $>3\sigma$ flux data points in the PDCSAP light curve with transits, eclipses, and the photometric modulation removed. We also adopt the flare rate $R_{31.5}$ from \cite{medina2020} where flares above an energy of $E=3.16\times 10^{31}$ ergs ($\log_{10}{E}=31.5$) are considered. Unlike in \cite{medina2020} and other works, we do not invoke a sophisticated Gaussian process (GP) model to remove systematic effects from the light curve while preserving the flare events. Since we do not use a GP, we are certainly not accounting for a number of flares, so our estimate of the flare rate $R_{31.5}$ of TOI-2119 will be a lower limit of $\log{R_{31.5}} \geq -0.70 \pm 0.32$ flares per day.} The strength and frequency of the flares indicate that this star is relatively active. \textcolor{black}{In Figure \ref{fig:periodogram} we show another Lomb-Scargle periodogram analysis of the normalised \textit{TESS} light curve with the flares retained, but the transits and eclipses of the BD removed. We see no strong evidence of any periodicity that would correlate the timing of the flares with the photometric modulation, transits, or eclipses.}

\begin{table}
\centering
 \caption[]{Coordinates and magnitudes for TOI-2119 (TIC 236387002). The $B_T$, $V_T$, $J$, $H$, $K$, WISE1, WISE2, and WISE3 values here are used to model the spectral energy distributions and constrain $T_{\rm eff}$ for the star.} \label{tab:toi_obs}
 \begin{tabular}{lccc}
 \hline
 {} & Description & Values  & Source\\
 $\alpha_{\rm J2000}$ &Equatorial& 16 17 43.17 & 1\\[2pt]
 $\delta_{\rm J2000}$ &coordinates& 26 18 15.16 & 1\\[2pt]
 $d$ & Distance (pc)& $31.817 \pm 0.026$ & 1\\[2pt]
 $T$\dotfill & \tess\, $T$\dotfill & $10.398 \pm 0.007$ & 2\\[2pt]
 $G$\dotfill & Gaia $G$\dotfill & $11.473 \pm 0.001$ & 1\\[2pt]
 $B_T$\dotfill & Tycho $B_T$\dotfill & $13.864 \pm 0.044$ & 3\\[2pt]
 $V_T$\dotfill & Tycho $V_T$\dotfill & $12.390 \pm 0.057$ & 3\\[2pt]
 $J$\dotfill & 2MASS $J$\dotfill & $8.976 \pm 0.019$ & 4\\[2pt]
 $H$\dotfill & 2MASS $H$\dotfill & $8.393 \pm 0.033$ & 4\\[2pt]
 $K_S$\dotfill & 2MASS $K_S$\dotfill & $8.139 \pm 0.021$ & 4\\[2pt]
 WISE1\dotfill & WISE 3.4$\rm \mu m$\dotfill &  $8.045 \pm 0.030$ & 5\\[2pt]
 WISE2\dotfill & WISE 4.6$\rm \mu m$\dotfill & $7.964 \pm 0.030$ & 5\\[2pt]
 WISE3\dotfill & WISE 12$\rm \mu m$\dotfill & $7.876 \pm 0.030$ & 5\\[2pt]
 WISE4\dotfill & WISE 22$\rm \mu m$\dotfill & $7.777 \pm 0.150$ & 5\\[2pt]
 \hline
 \end{tabular}
 \begin{list}{}{}
 \item[References:] 1 - \cite{edr3}, 2 - \cite{stassun18}, 3 - \cite{tycho2}, 4 - \cite{2MASS}, 5 - \cite{WISE}
 \end{list}
\end{table}

\subsection{Ground-based followup photometry}
We have additional seeing-limited transit photometry of TOI-2119 from \textcolor{black}{three} ground-based facilities: the Las Cumbres Observatory Global Telescope \citep[LCOGT;][]{Brown:2013} 1.0-m network node at McDonald Observatory (LCO-McD), the Multicolor Simultaneous Camera for studying Atmospheres of Transiting exoplanets instrument (MuSCAT2) \citep{narita2019} on the 1.52-m Carlos Sanchez Telescope, \textcolor{black}{and the Michigan State University (MSU) 0.6-m telescope.} 

\begin{figure}
\centering
\includegraphics[width=0.45\textwidth, trim={0.5cm 0.0cm 0.5cm 0.0cm}]{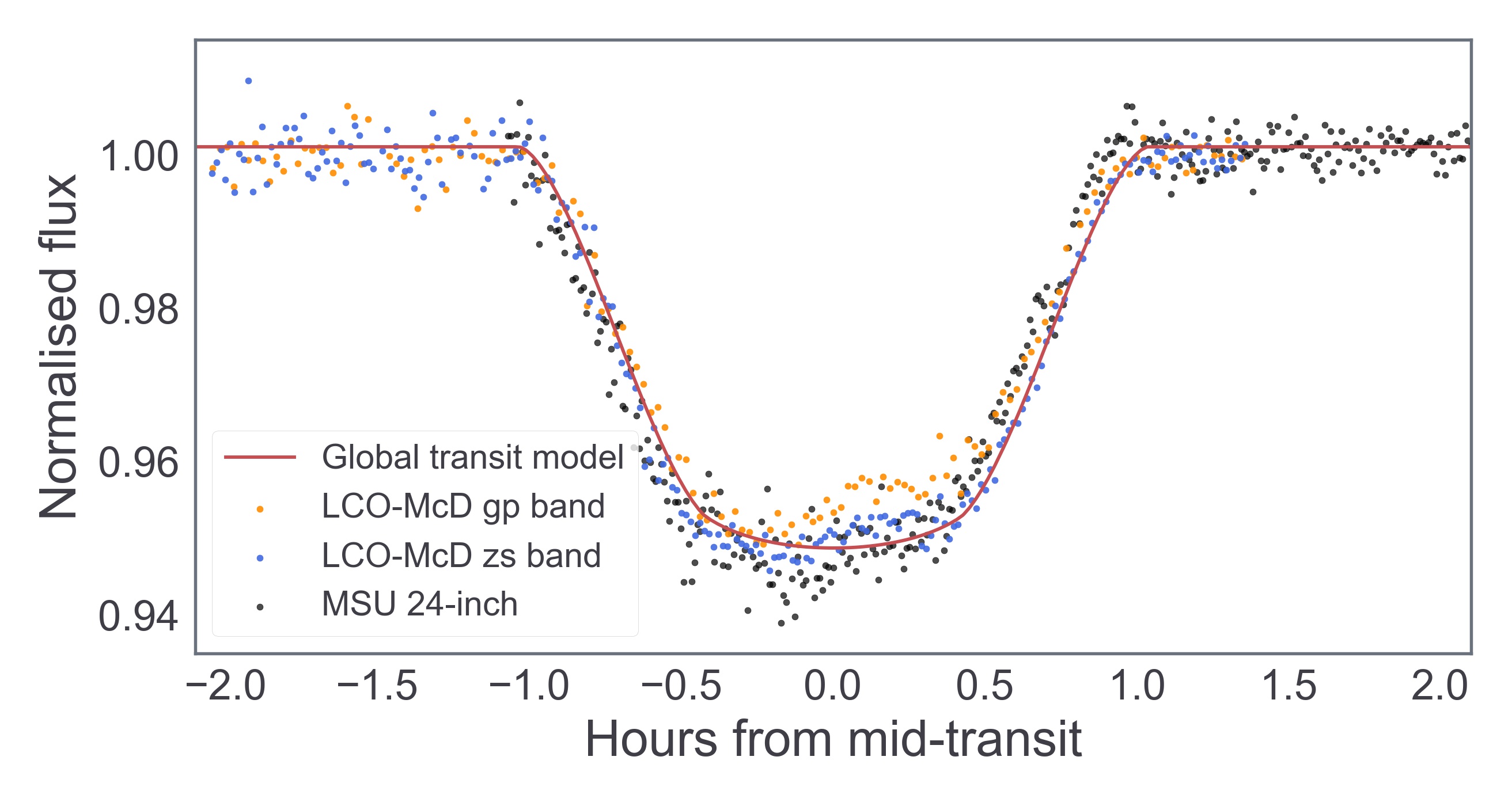}
\caption{Transit data from LCO-McD \textcolor{black}{and the MSU 0.6-m telescope}. These data are not used to model the transit (shown in red), but we present them here to show the extent to which the transit depth \textcolor{black}{changes during star spot crossings.} The star spot crossing has a greater effect on the transit depth in the LCO-McD $g_p$ bandpass than the $z_s$ bandpass. \textcolor{black}{Since the MSU data are taken at a different time than the LCO-McD (11 months apart), we have some sense of the spot evolution and can see how the effect from the spots almost vanish in the MSU data. Also note that the transit timing for the MSU data appears to be slightly earlier than the LCO data taken a year before. This difference is only on the order of minutes.}}\label{fig:lco}
\end{figure}

\textcolor{black}{The MSU 0.6-m telescope is equipped with $1024\times 1024$ Apogee ALTA U47 CCD camera providing $9.4\arcmin \times 9.4\arcmin$ field of view at a scale of $0\farcs55$ per pixel. The unfiltered images of TOI-2119 were obtained on 2022 March 4. We used the VaST code \citep{2018A&C....22...28S} for dark frame subtraction and flat-fielding and AstroImageJ for photometry and detrending.} The LCOGT observations were taken during the night of 2021 April 21 and cover one full transit of TOI-2119b. The 1-m telescope is equipped with $4096\times4096$ SINISTRO cameras having an image scale of $0\farcs389$ per pixel, resulting in a $26\arcmin\times26\arcmin$ field of view. The images were calibrated by the standard LCOGT {\tt BANZAI} pipeline \citep{McCully:2018}, and photometric data were extracted using {\tt AstroImageJ} \citep{Collins:2017}. We see noticeable asymmetry during the LCO-McD transit that we attribute to star spot crossings (Figure \ref{fig:lco}) \textcolor{black}{and much smaller spot crossings in the MSU 0.6-m data taken nearly 1 year later on 8 March 2022}. We use these follow up observations to confirm that the transit is on the target star, but we do not include this in our transit analysis.

The MuSCAT2 observations were taken during the night of 2021 May 12 in 4 simultaneous bandpasses: Sloan $g^\prime$, $r^\prime$, $i^\prime$, and $z_s$. Exposure times were 25, 9, 15, and 15 s for $g^\prime$, $r^\prime$, $i^\prime$, and $z_s$, respectively. There was some in-transit asymmetry in these observations that we correct for in order to use these data in the transit analysis. The data from the different bandpasses was reduced using the procedure and pipeline described in \cite{parviainen2020}.

\subsection{SOAR speckle imaging}
\begin{figure}
\centering
\includegraphics[width=0.45\textwidth, trim={0.0cm 0.0cm 0.0cm 0.0cm}]{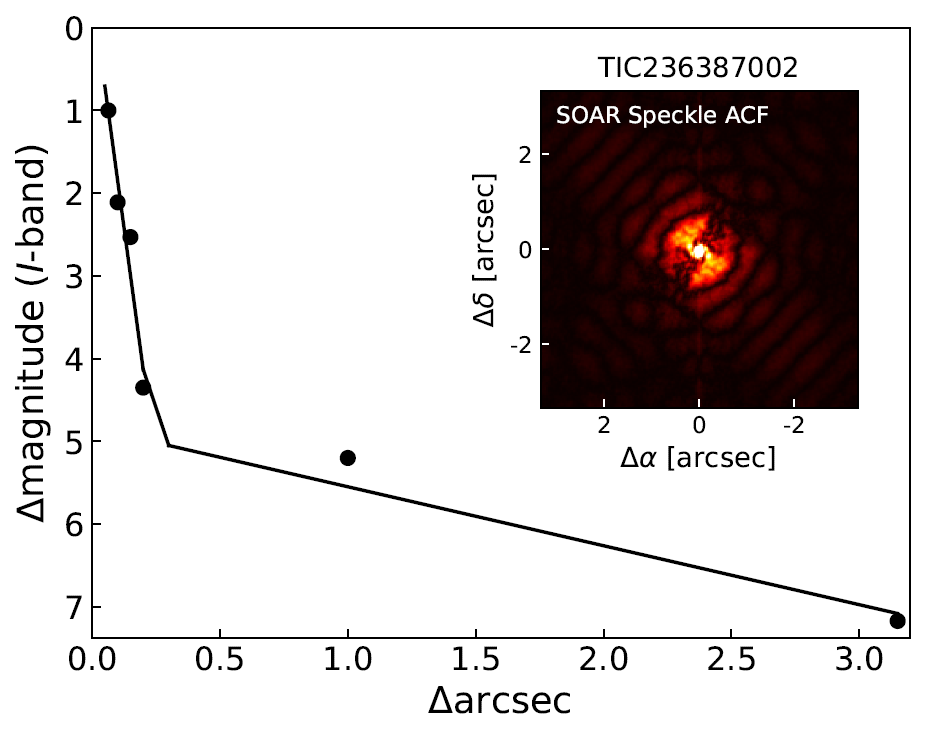}
\caption{The 5$\sigma$ sensitivity limits and auto-correlation functions of the SOAR speckle observations of TOI-2119. The black circles are measured data points and the lines are fits in different separation regimes. No nearby contaminating sources are detected within 3$\arcsec$.}\label{fig:soar}
\end{figure}

On 2021 April 25, we took SOAR speckle observations of TOI-2119 in the Cousin-\textit{I}-band, which is a similar bandpass to that of \tess. Further details of how SOAR observations are carried out are available in \cite{ziegler2019_soar}. A 5$\sigma$ detection speckle auto-correlation functions from the observations are shown in Figure \ref{fig:soar}. No nearby stars were detected within 3$\arcsec$ of TOI-2119 in the SOAR observations. We performed a search using data from Gaia EDR3 and find no stars brighter than $G = 15.2$ within \textcolor{black}{$65\arcsec$} of TOI-2119, \textcolor{black}{but given the renormalised unit weight error \citep[RUWE;][]{ruweDR2, edr3_ruwe} value is 1.9, an unresolved stellar companion cannot be ruled out, even with speckle imaging with SOAR. The RUWE is a measure of how confident an object is known to be a single source, where values that deviate from 1 are less likely to be single sources.}

\subsection{TRES spectra}\label{sec:tres}
We used the TRES instrument on Mt. Hopkins in Arizona, USA to obtain 8 follow up spectra for TOI-2119. TRES has a resolving power of $R \approx 44\,000$ and covers a wavelength range of 3900\AA\, to 9100\AA. We use 14 echelle orders between this wavelength range for each spectrum to measure a relative RV for each. We visually review each order to omit those with low signal-to-noise per resolution element (S/N) and to remove cosmic rays. In the case of this M-dwarf, we omitted the orders at bluer wavelengths shorter than 4800\AA. Each order is cross-correlated with the highest observed S/N spectrum of the target star and then the average RV of all the orders per spectrum is taken as the RV of the star for that observation. The exposure times for these follow up spectra range from 1200s to 2800s to give a S/N range of 9.1 to 20.7. 

\begin{figure}
\centering
\includegraphics[width=0.48\textwidth, trim={0.0cm 0.0cm 0.0cm 0.0cm}]{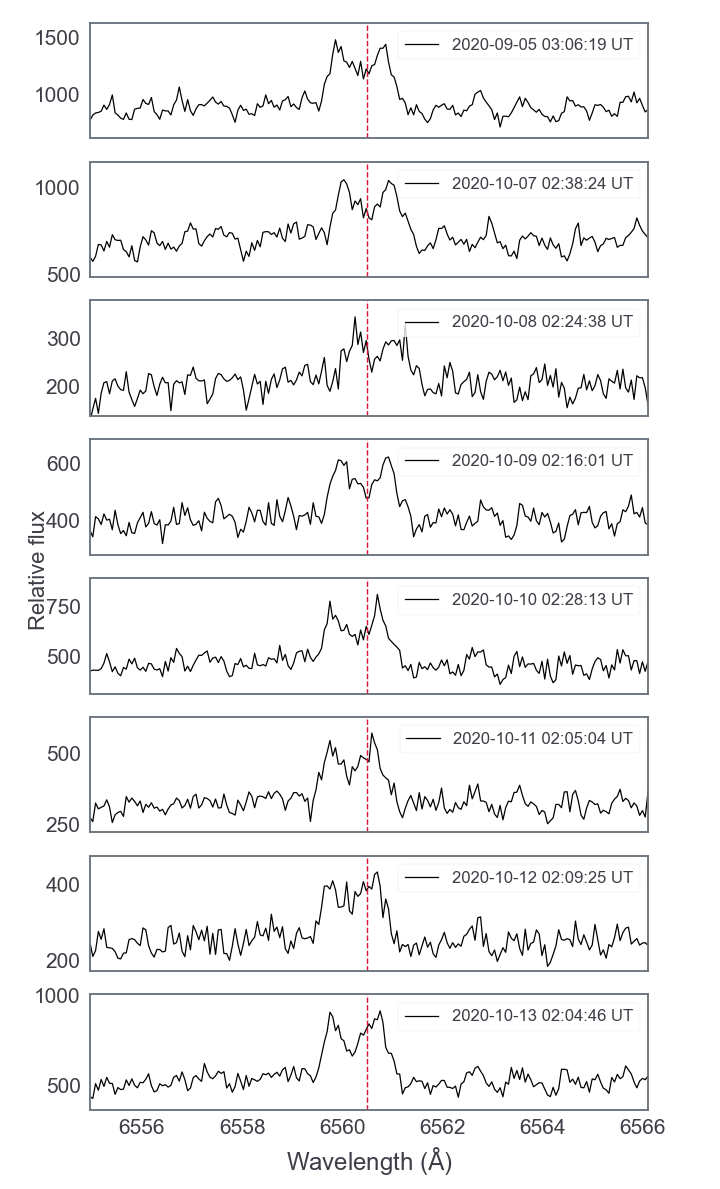}
\caption{H$\alpha$ feature from each of the TRES spectra for TOI-2119. The red vertical line serves as a visual aid for the relative shift in the self-absorbed line core at different times. The spectra have all been uniformly shifted to an arbitrary offset near the H$\alpha$ feature. The characteristic `double-peaked' shape is a result of optically-thick chromospheric scattering of H$\alpha$. We do not find any significant correlation with the strength of the H$\alpha$ feature or the symmetry of the peaks on either side of the centroid with the orbital phase of the BD's orbit.}\label{fig:halpha}
\end{figure}

We use the stellar parameter classification (SPC) software package by \cite{spc} to derive the projected stellar equatorial velocity $v\sin{I_\star}$ from co-added TRES spectra of TOI-2119. SPC uses a library of calculated spectra in the $\rm 5030-5320$\AA\, wavelength range, centered near the Mg b triplet. We report an upper limit on the projected rotational velocity for the star of $v\sin{I_\star} \leq 3.04 \pm 0.5$ $\rm km\, s^{-1}$. This $v\sin{I_\star}$ measurement is relatively low and potentially becomes degenerate with other spectroscopic broadening features of the stellar photosphere that do not indicate the true projected rotational velocity of the star.

As SPC is not suited to accurate spectral analysis of low-mass stars like TOI-2119, we instead use empirical relations for early-type M-dwarf metallicities \citep{mann2013} and $T_{\rm eff}$ \citep{mann2015} to estimate these values for TOI-2119. These relations rely on $J-K$ magnitude values that we obtain from the TESS Input Catalogue v8.2 \citep{stassun2019TIC} and they yield an $\rm [Fe/H] =-0.10 \pm 0.09$ and $T_{\rm eff} = 3512 \pm 100$K. We use these values as priors in our analysis of the stellar parameters.

\begin{table}
    \centering
    \setlength{\tabcolsep}{3pt}
    \caption[]{Relative radial velocities of TOI-2119 from TRES. The signal-to-noise per resolution element (S/N) is listed in the fourth column. The spectrum with the highest S/N is chosen as the zero-point.} \label{tab:rvs}
    \begin{tabular}{ccccc}
    \hline
$\rm BJD_{\rm TDB}-2450000$ & RV ($\rm m\, s^{-1}$) & $\sigma_{\rm RV}$ ($\rm m\, s^{-1}$) & S/N & Exp. Time (s)\\
9097.645269 &   0.0  &   70.0 &  20.7  & 2800\\[2pt]
9129.623800 &   12655.3  &  65.5 &  17.9 & 2800\\[2pt]
9130.604995 &   20381.2  &  106.0 &  9.1  & 1200\\[2pt]
9131.599229 &   9252.9  &   50.98 &  13.6  & 1300\\[2pt]
9132.607167 &   2770.6 &    51.0 &  14.3  & 1200\\[2pt]
9133.591252 &   519.05  &  122.0 &  10.9  & 1200\\[2pt]
9134.593961 &   590.6  &   99.5 &  10.2  & 1200\\[2pt]
9135.599268 &   3242.9  &   70.0 &  15.8  & 2700\\[2pt]
\hline
    \end{tabular}
\end{table}

\subsubsection{Double-peaked H$\alpha$ feature}
One curious characteristic of these TRES spectra is seen in the H$\alpha$ line profile as a `double-peaked' or `horned' feature. Figure \ref{fig:halpha} shows the H$\alpha$ line for TOI-2119 and this `horned' shape is typically attributed to the presence of a circumstellar disc. However, given the relatively small width of the profile, we attribute this to self-absorption (or self-reversal) of H$\alpha$ instead of the presence of a disc. This self-absorption occurs due to optically-thick scattering in the chromosphere of the star \citep[see][for more details]{youngblood2022}. The symmetry (relative height difference between each peak) of the horned H$\alpha$ does not correlate with the orbital period ($P_{\rm orb}=7.2$ days) of the transiting BD or the rotation period ($P_{\rm rot}=13.11$ days) of the star. We do not have simultaneous photometry with these TRES spectra to check whether or not the horned feature is linked to a flare event, so we cannot confidently connect the two. Although the presence of emission of H$\alpha$ does indicate some activity from TOI-2119, the relatively smaller equivalent width (EW) of H$\alpha$ of roughly $\sim$1\AA\, is a sign that this star is only moderately active.

\textcolor{black}{We also consider the possibility that this peaked feature, which is also present in the sodium doublet, may instead be from an unresolved stellar companion. The TRES pixel scale at the H$\alpha$ feature is 1 pixel per 0.0557$\angstrom$, or 45.5$\rm km\,s^{-1}$ per $\angstrom$. The separation between the peaks seen in the H$\alpha$ profile ranges from roughly $0.9-1.1\angstrom$ and this translates to a range of plausible orbital separations, eccentricities, and masses for an unseen stellar companion such that this spectral double-peaked feature is produced but also that the companion evades detection by SOAR. The ranges of possible orbital separations, eccentricities, and stellar companion masses are roughly $1-2$ AU, $e>0.5$, and $100-400\mj$, respectively. However, this does not enable us to confirm that such a companion exists given the self-absorption behavior M-dwarfs exhibit not only for H$\alpha$, but for other features that may show a similar separation. Long-term RV observations of the primary star are necessary to narrow the range of plausible values and to see a change in the wavelength differences between the peaks in H$\alpha$.}

\begin{figure}
\centering
\includegraphics[width=0.48\textwidth, trim={0.0cm 0.5cm 0.0cm 0.0cm}]{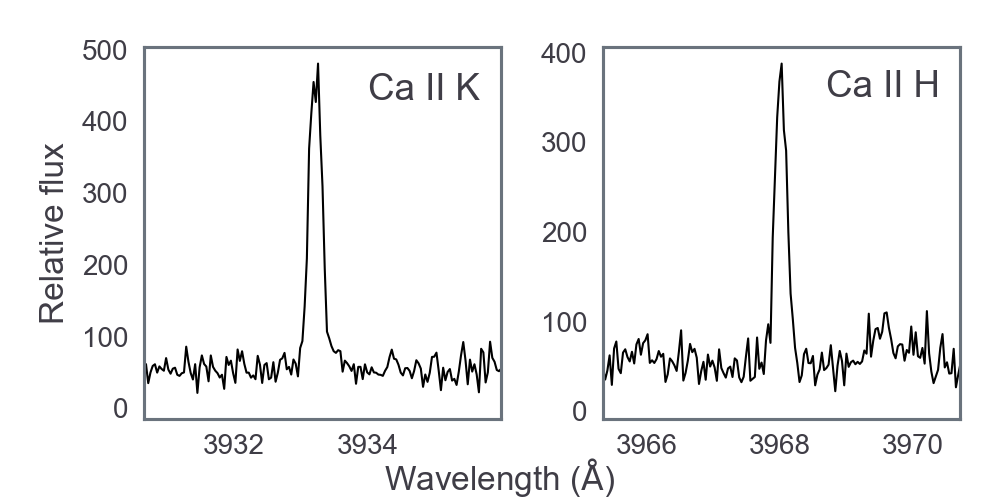}
\caption{Ca II H and Ca II K line profile of TOI-2119. The strength of this line indicates that this star is relatively active. Here, we show these profiles from a coadded spectrum of all the TRES observations of TOI-2119. \textcolor{black}{We do not detect enough continuum flux (in arbitrary units here) to perform a reliable S-index measurement.}}\label{fig:caII_HK}
\end{figure}

We also find Ca II H and Ca II K emission from the spectra of TOI-2119 (Figure \ref{fig:caII_HK}), \textcolor{black}{and although this provides some  qualitative evidence for the star's activity, we do not have enough continuum flux around the Ca II H/K features to perform a proper S-index derivation using the techniques described in \cite{mayo2018} specifically for TRES.} There are also no known published values for the S-index for this system that we can refer to here. \textcolor{black}{Lastly, we do not detect an appreciable ``double-peaked'' behavior to the Ca II H/K features, which may be due to the fact that the self-absorption phenomenon lessens in intensity at lower stellar masses for certain features, including Mg II and Ca II \citep{wood2005}.}


\section{Analysis}\label{sec:analysis}

\subsection{Global analysis with {\tt EXOFASTv2}}\label{sec:exofast}
Here we give details on our global analysis of the stellar and BD parameters for TOI-2119 using {\tt EXOFASTv2} \citep{eastman2019}. {\tt EXOFASTv2} uses the Monte Carlo-Markov Chain (MCMC) method and for each MCMC fit, we use N=36 (N = 2$\times n_{\rm parameters}$) walkers, or chains, and run until the fit passes the default convergence criteria for {\tt EXOFASTv2} \citep[described in][]{eastman2019}.

\begin{figure}
\centering
\includegraphics[width=0.48\textwidth, trim={0.0cm 0.0cm 0.0cm 0.0cm}]{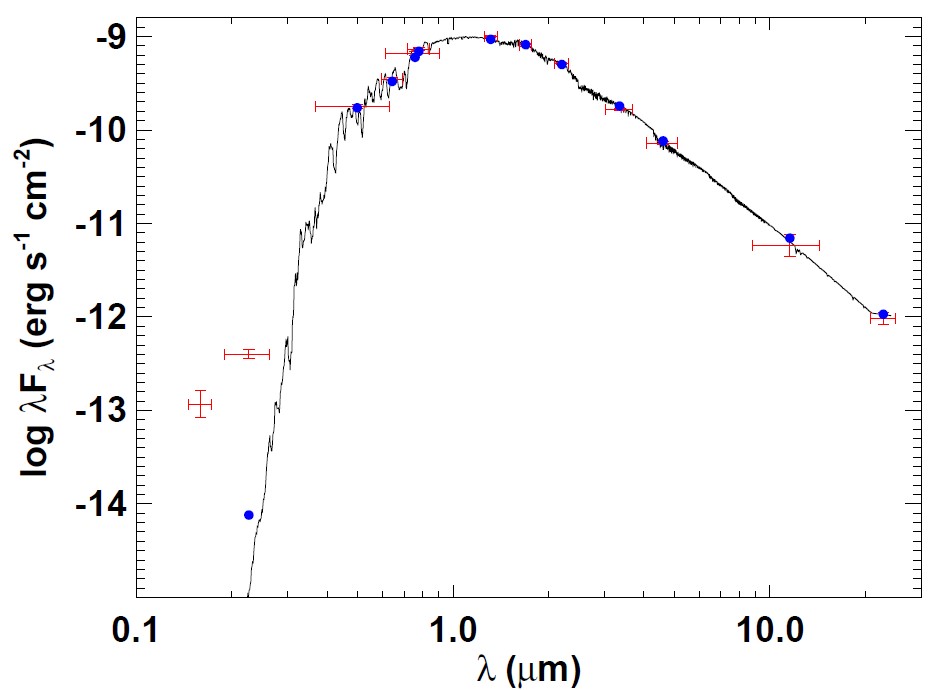}
\caption{{\tt EXOFASTv2} SED for TOI-2119. The best-fitting value of $T_{\rm eff}$ derived here is $T_{\rm eff} = 3624\pm 38$K. The red symbols represent the observed photometric measurements, where the horizontal bars represent the effective width of the bandpass. Blue symbols are the model fluxes from the best-fitting Kurucz atmosphere model (black). We do not fit the data in the UV bandpasses (around $0.2-0.3\rm \mu m$) given the significant UV excess from the star, which is likely linked to the flares (Figure \ref{fig:tess_lc}) and the star's apparent activity.}\label{fig:sed}
\end{figure}

Here we will describe our inputs into {\tt EXOFASTv2} and what parameters we obtain from each one. First, we input the stellar magnitudes referenced in Table \ref{tab:toi_obs} and use the $T_{\rm eff}$ empirical relations from \cite{mann2015} as a prior on the SED model for the star (Figure \ref{fig:sed}). Note the UV excess in the SED for TOI-2119. We attribute this to stellar activity, likely related to the strength of flares that we see in the \tess\, light curve for the star. \textcolor{black}{The broadband UV data points from GALEX \citep[][]{bianchi2017_galex} cannot be used to directly estimate $\log{R^\prime_{HK}}$ following the technique used in \cite{findeisen2011} due to the $B-V$ magnitude color of TOI-2119 falling outside of the valid range determined in that study.}

We also input parallax measurements from Gaia EDR3 ($\varpi = 31.77 \pm 0.03$ mas), which are used with the SED model and an upper limit on V-band extinction \citep[$A_V \leq 0.05$,][]{av_priors} to determine the stellar luminosity and radius. We use this stellar radius with the radius ratios obtained from our input \tess\, transit photometry to constrain the radius of the transiting BD. The inclination from the transit data are combined with our input RV follow up from TRES to constrain the mass and orbital properties of TOI-2119b. The relative RVs are done with respect to the target star, so the zero-point is arbitrary (Table \ref{tab:rvs}). We let the RV offset $\gamma$ be a free parameter and we include an RV jitter term, $\sigma_j$, to account for the surface activity of the star. The RV jitter term derives from a white noise model implemented in {\tt EXOFASTv2}. Our input $\rm [Fe/H] =-0.10 \pm 0.09$ and $T_{\rm eff} = 3512 \pm 100$K empirical estimates from \cite{mann2013} and \cite{mann2015}, respectively, are used as priors on the build-in MIST stellar isochrone models \citep{mist1, mist2, mist3} in {\tt EXOFASTv2}. These models are detailed in these three studies, but we note here that the MIST models are less accurate in determining the ages of low-mass stars like TOI-2119, so our priors on the stellar mass, radius, and age are especially relevant here.

\begin{figure}
    \centering
    \includegraphics[width=0.45\textwidth, trim={1.0cm 0.0cm 1.0cm 0.0cm}]{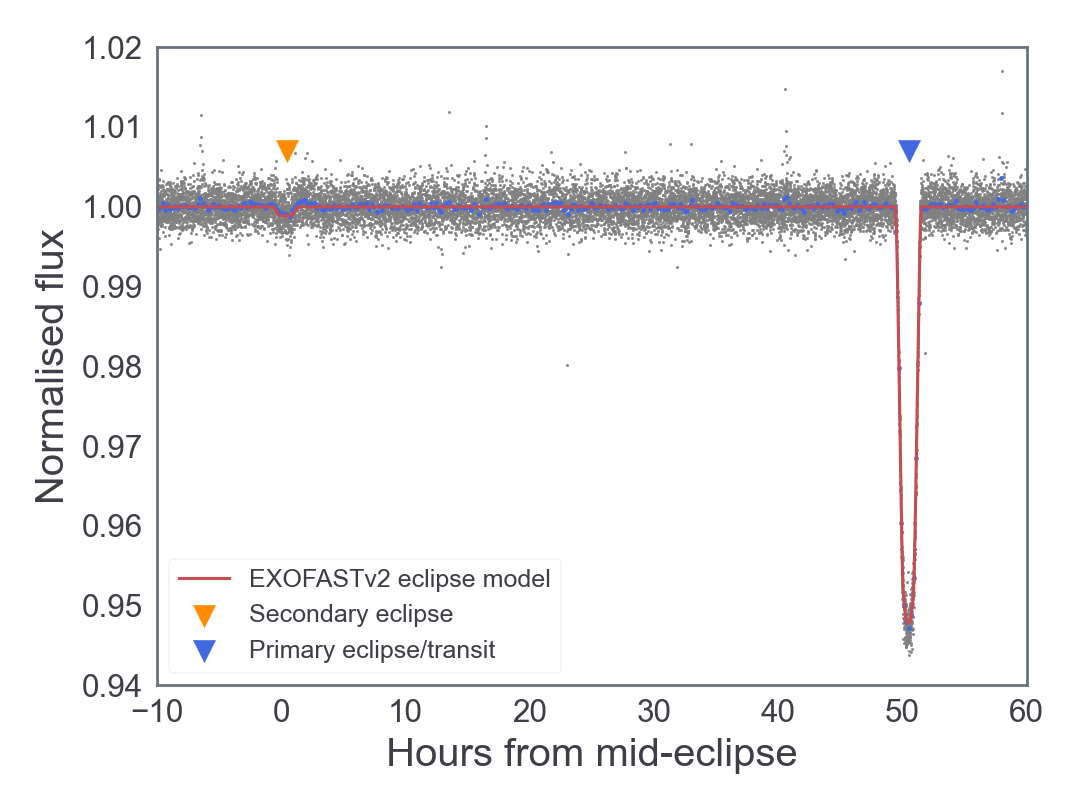}
    \caption{Normalised \tess\, light curve phased to show primary transit and secondary eclipse.}
    \label{fig:eclipse_transit}
\end{figure}

\begin{figure}
\centering
\includegraphics[width=0.47\textwidth, trim={0.0cm 0.0cm 0.0cm 0.0cm}]{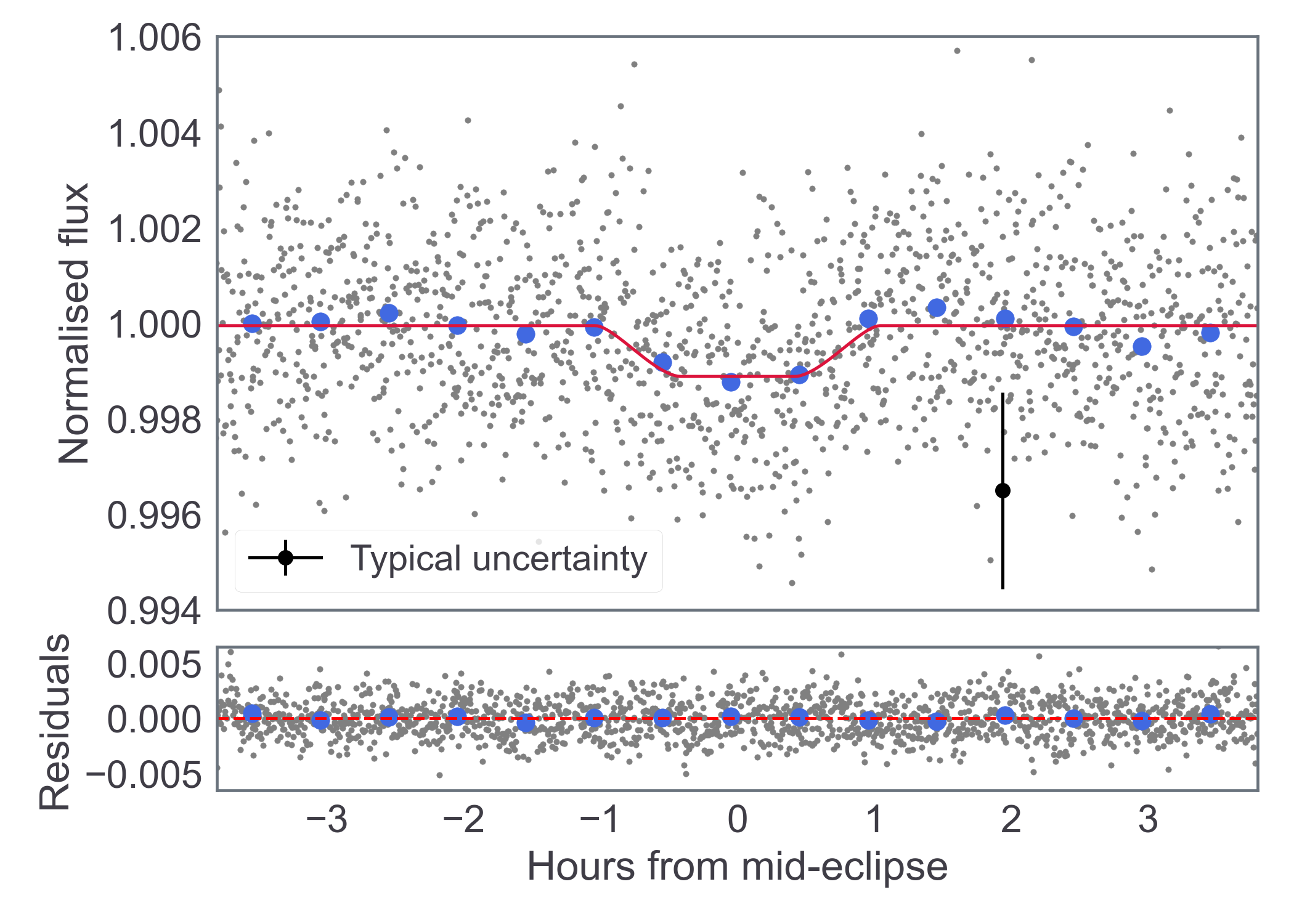}
\includegraphics[width=0.47\textwidth, trim={0.0cm 0.0cm 0.0cm 0.0cm}]{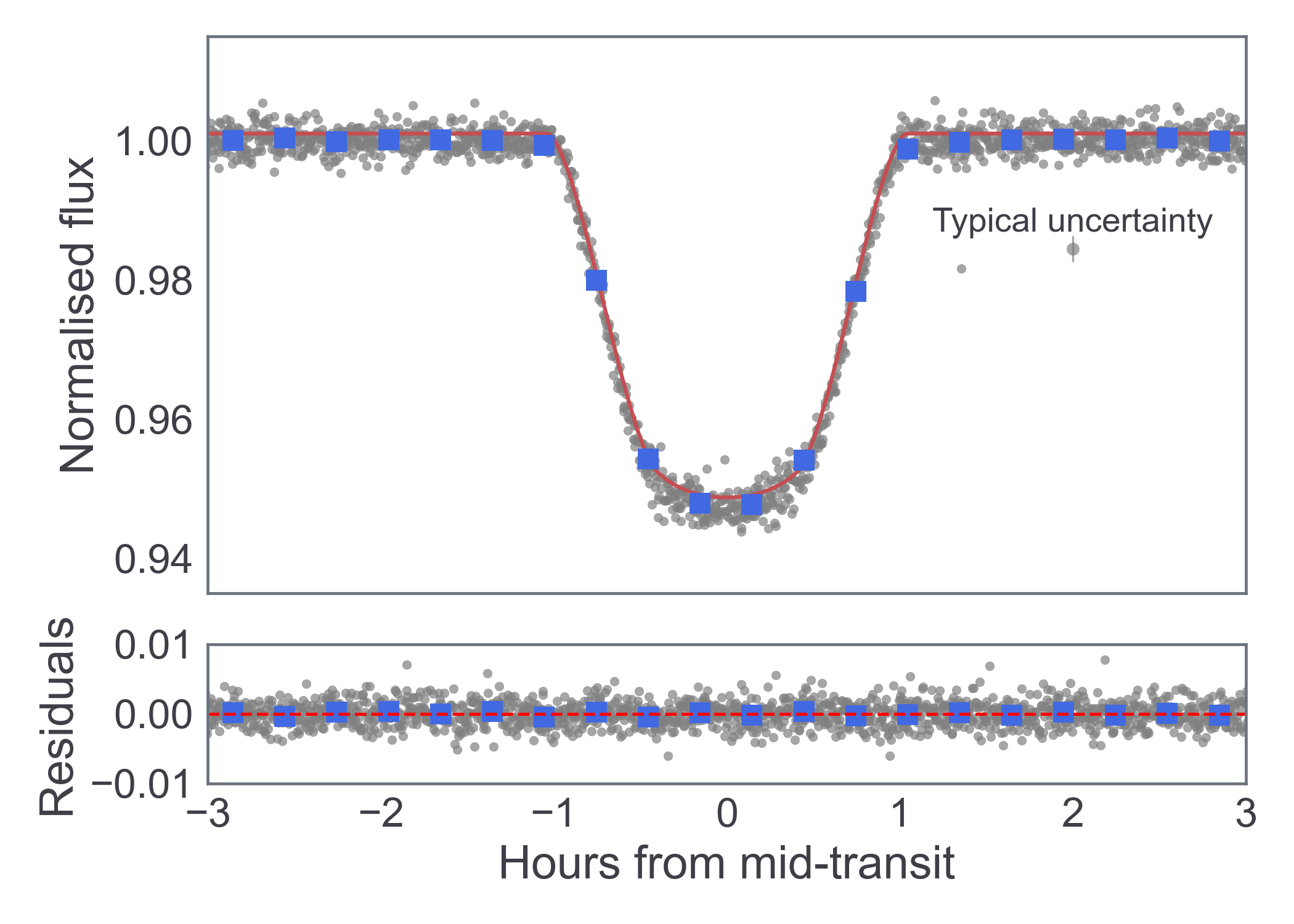}    
\includegraphics[width=0.47\textwidth, trim={0.0cm 0.0cm 0.0cm 0.0cm}]{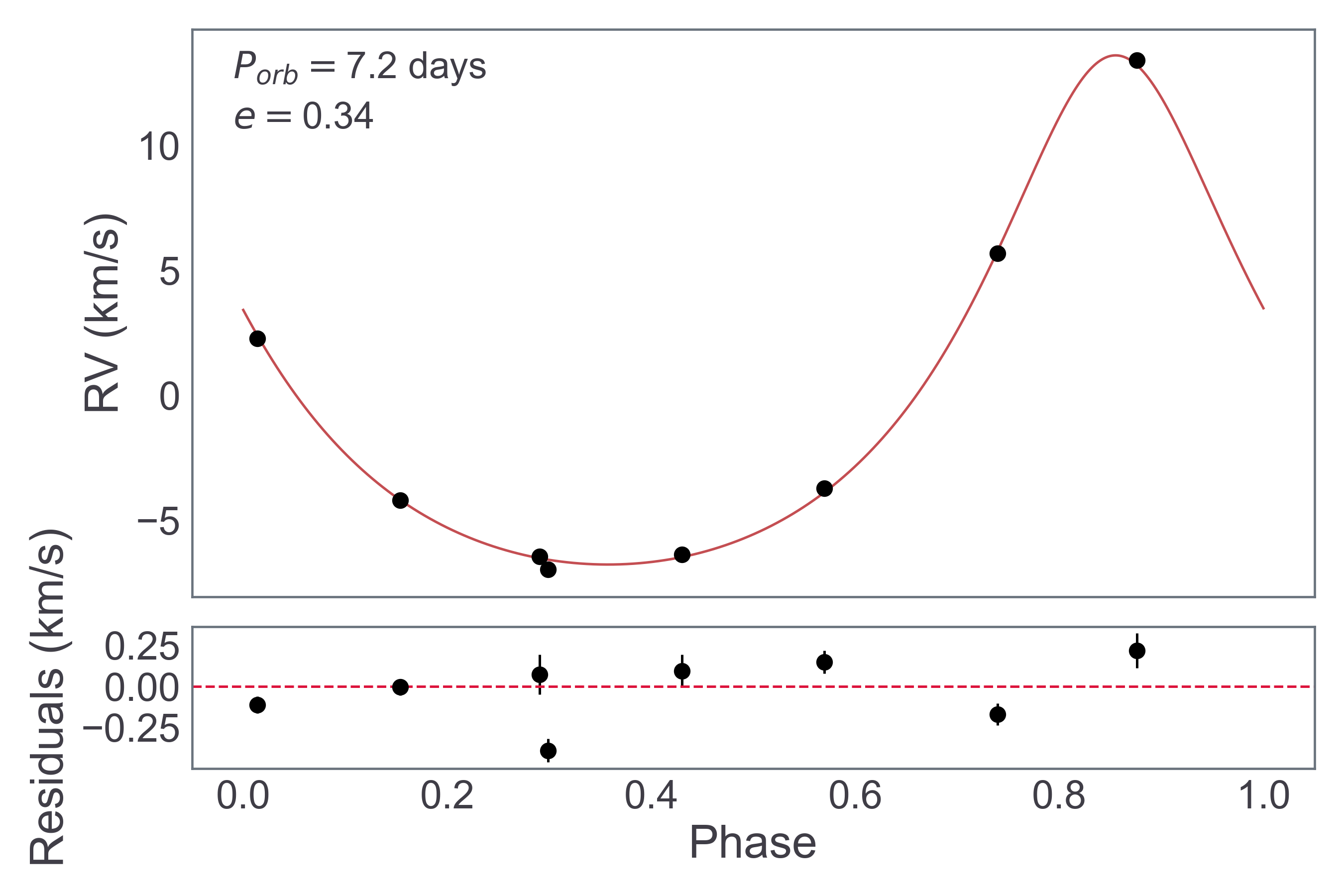}
\caption{Top: \tess\, eclipse light curve of TOI-2119 with {\tt EXOFASTv2} eclipse model in red. Binned data are in blue. The eclipse of the BD is roughly 0.1\% (1000 ppm) deep. Middle: \tess\, transit light curve of TOI-2119 with {\tt EXOFASTv2} transit model in red. Bottom: TRES multi-order relative radial velocities of TOI-2119 with {\tt EXOFASTv2} orbital solution plotted in red.}\label{fig:toi2119_obs}
\end{figure}

\begin{figure}
    \centering
    \includegraphics[width=0.40\textwidth, trim={1.0cm 0.0cm 1.0cm 0.0cm}]{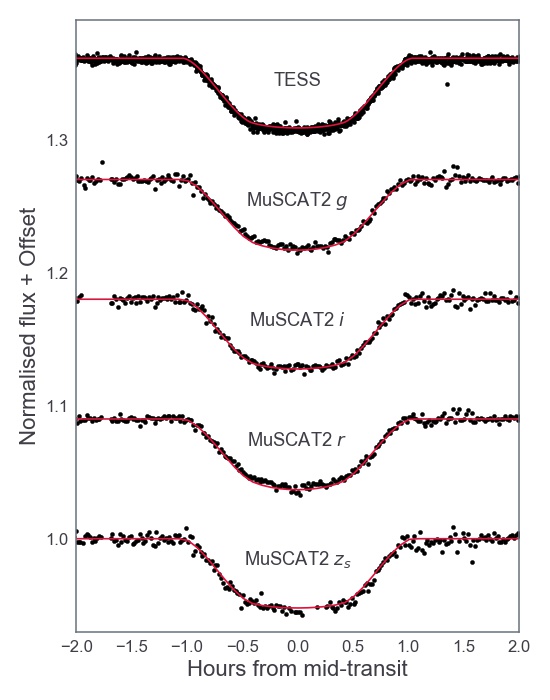}
    \caption{\tess\, and MuSCAT2 light curves for TOI-2119. There was a cloud passage event during the transit in the MuSCAT2 data and this appeared as a large spike in all four bandpasses. The detrending algorithm was able to suppress the spike in all bandpasses except the $z_s$ one, so this required the removal of some in-transit data in that bandpass.}
    \label{fig:muscat2}
\end{figure}

We set uniform $\mathcal{U}[a,b]$ or Gaussian $\mathcal{G}[a,b]$ priors on our input parameters. We use parallax measurements from Gaia EDR3 and the M-dwarf empirical relations for mass, radius, age, $T_{\rm eff}$, and [Fe/H] from \cite{mann2015} and \cite{engle2018} (both described in the next section) to define our Gaussian priors, which have width $b$ and mean $a$ of the each parameter. \cite{eastman2019} gives a detailed description of how priors are implemented in {\tt EXOFASTv2}. The full set of derived parameters and input priors for the system is shown in Tables \ref{tab:muscat} and \ref{tab:exofast_toi2119}. The SED derived by {\tt EXOFASTv2} for the star are shown in Figure \ref{fig:sed}. The orbital solution and transit model for TOI-2119 is shown in Figure \ref{fig:toi2119_obs} with the additional light curves from MuSCAT2 shown in Figure \ref{fig:muscat2}. We find marginal differences in the transit depths between the the different bandpasses for the light curve data.

\subsubsection{Determining the properties of the M-dwarf}
We rely on the \cite{mann2015} relations and the \cite{engle2018} relations as aids in our analysis of the stellar mass, radius, and age for TOI-2119. The \cite{mann2015} relations are constructed from empirical data of M-dwarf absolute magnitude-mass correlations and absolute magnitude-radius correlations from a sample of 183 stars ranging from M7 to K7 type stars. For TOI-2119, we use the following absolute K-band relations from \cite{mann2015} to estimate a stellar mass and radius:

\begin{equation}\label{eq:mstar}
    M_{\star} = a_1 + b_1K + c_1K^{2} + d_1K^{3} + e_1K^{4}
\end{equation}

\noindent where $a_1=0.5858$, $b_1=0.3872$, $c_1=-0.1217$, $d_1=0.0106$, $e_1=-2.7262\times 10^{-4}$, and $K$ is the absolute $K_S$ magnitude, which uses $K_S$ from Table \ref{tab:toi_obs} in the equation $K = K_S + 5\log{d} +1$ where $d$ is the distance to TOI-2119 in kiloparsecs.

\begin{equation}\label{eq:rstar}
    R_{\star} = a_2 + b_2K + c_2K^{2}
\end{equation}

\noindent where $a_2=1.9515$, $b_2=-0.3520$, and $c_2=0.01680$. $K$ is again the absolute $K_S$ band magnitude. Using Equations \ref{eq:mstar} and \ref{eq:rstar}, we find $M_\star = 0.522 \pm 0.009\mst$ and $R_\star = 0.499 \pm 0.014\rst$. The reported uncertainties here are percentage values of $M_\star$ and $R_\star$, following the procedure from \cite{mann2015}. We apply $M_\star$ and $R_\star$ as priors in our {\tt EXOFASTv2} analysis of the TOI-2119 system.

\begin{table*}
\centering
    \caption[]{Limb-darkening coefficients and individual filter transit depths for MuSCAT2 and \tess\, observations (Figure \ref{fig:muscat2}).} \label{tab:muscat}
    \begin{tabular}{cccccc}
    \hline
{} & Sloan $g^\prime$ & Sloan $r^\prime$ & Sloan $i^\prime$ & Sloan $z_s$ & \tess\, \\
$u_{1}$ &   $0.485\pm0.041$  &   $0.510\pm0.045$  &  $0.318\pm 0.042$  & $0.143\pm 0.044$ & $0.288\pm 0.046$\\
$u_{2}$ &   $0.277\pm0.047$  &  $0.251\pm 0.053$ &  $0.306\pm 0.046$ & $0.376\pm0.046$ & $0.355\pm 0.052$\\
$\delta$ & $0.048\pm 0.001$ & $0.048\pm 0.001$ & $0.048 \pm 0.001$ & $0.049 \pm 0.001$& $0.049\pm 0.006$\\
    \hline
    \end{tabular}
\end{table*}

To better estimate the age of TOI-2119, we use the relationship between the rotation period and stellar age for M-dwarfs explored in \cite{engle2018}. In that study, \cite{engle2018} present empirical relations between the rotation periods and ages of early (M0--M1) and mid (M2.5--M6) M-dwarfs. In our case, we find that TOI-2119 is more likely an M0--M1 star based on the $T_{\rm eff} = 3512$K and stellar mass $M_\star = 0.53\mst$. The rotation-age relation for early-type M-dwarfs from \cite{engle2018} is: 

\begin{equation}\label{eq:age}
    t_\star = y_0 + aP_{\rm rot}^b
\end{equation}

\noindent where $t_\star$ is in Gyr, $y_0= 0.365 \pm 0.431$, $a = 0.019 \pm 0.018$, $b = 1.457 \pm 0.214$, and $P_{\rm rot} = 13.11 \pm 1.41$ days from our periodogram analysis of the \tess\, light curve. By using Equation \ref{eq:age}, we find an approximate age estimate of TOI-2119 to be $t_\star = 1.17 \pm 1.15$ Gyr.

We also search for TOI-2119 with the {\tt BANYAN $\rm \Sigma$} \citep{gagne2018} tool to check for association with moving groups and clusters of stars with known ages. \textcolor{black}{We use the 6D kinematic information available for TOI-2119 ($\alpha_{\rm J2000}$, $\delta_{\rm J2000}$, $\varpi$, proper motion, and the Gaia DR2 radial velocity) and find that TOI-2119 has a >99\% probability of belonging to the field, not a part of a stellar cluster or moving group.}

\subsubsection{Different scenarios for the stellar age}\label{subsec:ages}
Given the importance of a relatively precise determination of the stellar age to testing substellar evolutionary models in this work, we present 3 different scenarios for the age of TOI-2119:

\begin{itemize}
    \item \textbf{Scenario 1, age prior, early M-dwarf:} This is the scenario we adopt for this star. This assumes that we reliably know the rotation period of the star ($P_{\rm rot}=13.11$ days) from the \tess\, data and that TOI-2119 is an early-type (M0--M1) M-dwarf. Under these assumptions, we use Equation \ref{eq:age} from \cite{engle2018} to set a prior $\mathcal{G}[1.17,1.15]$ which yields an age estimate of $2.1^{+1.1}_{-0.9}$ Gyr from {\tt EXOFASTv2}.
    \item \textbf{Scenario 2, age prior, mid M-dwarf:} We run a separate analysis assuming that the host star is a mid-type M-dwarf. Under these assumptions, we use the corresponding equation from \cite{engle2018} to set a prior $\mathcal{G}[0.85,0.25]$ which yields an age estimate of $0.93 \pm 0.24$ Gyr from {\tt EXOFASTv2}.
    \item \textbf{Scenario 3, $P_{\rm rot}$ from spectroscopic $v\sin{I_\star}$:} If we take the spectroscopic $v\sin{I_\star} = 3.04$ $\rm kms^{-1}$ and use $R_\star = 0.5\rst$, we can estimate a stellar rotation period of 8.3 days (this hold under the assumption $\sin{i} \approx \sin{I_\star}$), which places the age estimate for the star between $0.5-0.8$ Gyr, depending on whether it is assumed to be an early- or mid-type M-dwarf.
\end{itemize}

We run another scenario without any priors on the age of TOI-2119 and we see that {\tt EXOFASTv2} is uninformative on the age of the star in this case (meaning that the age ranges from $1-10$ Gyr at the 1$\sigma$ level). This means that the ages seen in our above scenarios are more reflective of the priors than an accurate position on the stellar isochrone tracks (from MIST) that {\tt EXOFASTv2} uses. With this in mind, we proceed by adopting the stellar ages directly from Equation \ref{eq:age} ($t_\star = 1.17 \pm 1.15$ Gyr).

We favor and adopt Scenario 1 because we have the necessary data from the light curve to estimate the rotation period of TOI-2119 and to determine that it is an early-type M-dwarf. However, we emphasize that if the spectroscopic classification of TOI-2119 has the star straddle the line between early- and mid-type M-dwarfs, then the M-dwarf may be much younger (Scenario 2). 

We also stress that using gyrochronology (age--rotation relationships) with M-dwarfs is challenging. This is due to a phenomenon known as `stalling', which is when a star does not slow in its rotation rate over time. \cite{curtis2020_gyro} find that this stalling of stellar spin-down becomes more prominent in lower mass stars and that stars of $M_\star=0.55\mst$ may stall until they are 1.3 Gyr in age.  Specifically, \cite{curtis2020_gyro} point out that slowly-rotating low-mass stars may be in a stalled state from 100 Myr to 1.3 Gyr (in the case of $M_\star\approx 0.55\mst$). With a mass of $M_\star=0.53\mst$ (see Table \ref{tab:exofast_toi2119}), TOI-2119 may be a star that has `stalled' in its rotation period and so its age may not yet be knowable to an precision better than $0.1-1.3$ Gyr (this fact is reflected in its age and age uncertainty are $t_\star = 1.17 \pm 1.15$ Gyr). This has important implications for using these age dating techniques for the star to test the substellar evolutionary models for the companion BD.

\subsection{Secondary eclipse and thermal emission from the brown dwarf}
The TOI-2119 system is a rare example of a detectable secondary eclipse of a BD. The eclipse impact parameter is $b_S=0.635$, so the BD is completely blocked by the M-dwarf along the observer's line of sight during the eclipse. The thermal emission of the BD may be characterised by the eclipse depth $\delta_{S}$, which is proportional to the radius ratio and the temperature ratio of the BD and star: $\delta_S \propto (R_{b}/R_\star)^2 \times (T_{b}/T_\star)$. The first detection of thermal emission from a hot Jupiter was published by \cite{charbonneau2005}, so we follow their method in presenting the thermal emission from TOI-2119b.

Using the \tess\, data, we report a secondary eclipse depth of $\delta_S=1053\pm 88$ parts per million (ppm) of units of relative flux in the \tess\, band. The \tess\, bandpass window, or spectral response function, is centered at roughly 800 nm (0.8 microns) and spans 600--1000 nm \citep{sullivan15}, so we are measuring the relative brightness of the BD to the star at only these wavelengths. We use this relative brightness to derive an effective temperature for the BD. Adopting the equation used for the secondary eclipse depth from \cite{charbonneau2005}, we have:

\begin{equation}\label{eq:emission1}
    \delta_S \approx \left(\frac{R_{b}}{R_\star}\right)^2 \frac{\int F_{b}(\lambda)S(\lambda)\lambda d\lambda}{\int F_\star(\lambda)S(\lambda)\lambda d\lambda}
\end{equation}

\noindent where $S(\lambda)$ is the \tess\, spectral response function, $F_b(\lambda)$ is the flux of the BD, $F_\star(\lambda)$ is the stellar flux, and the limits of the integrals are the lower and upper limits of $S(\lambda)$ (600--1000 nm) since it falls to zero outside of this bandpass wavelength range. In order to derive the BD effective temperature $T_{b, {\rm eff}}$, we simplify Equation \ref{eq:emission1} by assuming a blackbody profile for both the star and BD. \textcolor{black}{We note that the model atmospheres from \cite{sonora21} show several absorption features within the \tess\, bandpass especially strongly at temperatures below $T_{\rm eff} \leq 1500$K, so by choosing to implement a simpler blackbody profile for the BD, we are potentially ignoring the effect that these features have, even at higher temperatures where they are shallower. However, the \cite{sonora21} models are not designed to include the effects of stellar irradiation, and so, this further complicates a direct use of these BD atmospheric models in the case of the TOI-2119 system. Given this, we use a blackbody profile to represent the SED of the BD, acknowledging that despite the actual SED potentially being more feature-rich and complex, we do not yet have the full model framework or relevant data in other bandpasses to address the stellar irradiation that is present.} With that aside, our equation $F(\lambda) = \pi B(\lambda,T)$ where $B(\lambda,T)$ is the Planck function, changes Equation \ref{eq:emission1} to:

\begin{equation}\label{eq:emission2}
    \delta_S \approx \left(\frac{R_{b}}{R_\star}\right)^2 \frac{\int B_{b}(\lambda, T_{b,{\rm eff}})S(\lambda)\lambda d\lambda}{\int B_\star(\lambda,T_{\rm eff})S(\lambda)\lambda d\lambda}
\end{equation}

\textcolor{black}{From this, we find the maximum likelihood $\mathcal{L}$ (via the minimum $\chi^2(T)$) for $T_{b, {\rm eff}}$ to find that the BD has an effective blackbody temperature of $T_{b, {\rm eff}} = 2030 \pm 84$K.} 


\section{Discussion}\label{sec:conclusion}
Here we will examine what this latest addition to the transiting BD population contributes to our understanding of substellar evolutionary models from \cite{ATMO2020} (ATMO 2020) and \cite{sonora21} (S21 for `Sonora 2021'). However, before discussing the significance of TOI-2119b to the transiting BD population, let us first recapitulate our understanding of the nature of the host star.

\subsection{Host star activity}
We see evidence that TOI-2119 is a relatively active M-dwarf. This evidence for the activity is seen in the \tess\, data, the TRES spectra of the star, and the SED. 

The \tess\, data reveal numerous flares over the 60 days of observation that brighten to over $0.5-1.0$\% of the star's baseline brightness. Flares are known to be an indicator of magnetic activity for M-dwarfs \citep[e.g.][]{skumanich_1972, noyes1984} and are likely the cause of the UV excess seen in the SED (Figure \ref{fig:sed}) for this star. Studies like \cite{newton2017} and \cite{medina2020} (and references therein) have explored how the strength of H$\alpha$ and Ca II H/K emission are also indicators of stellar activity and so, we find it appropriate here to consider the strength of these features as a sign that TOI-2119 is at least somewhat magnetically active even though we cannot cite or derive an S-index value. \textcolor{black}{Regarding this, although we have a nominal detection of the Ca II H and K lines, we are not able to directly measure $\log{R^\prime _{HK}}$ to derive an S-index due to the relatively low continuum flux. This low continuum signal-to-noise around the line cores of the Ca II H/K feature means that we will overestimate an S-index when using the technique described in \cite{mayo2018} for TRES}.

We note that TOI-2119 is not representative of the most active M-dwarfs known to host BDs; a better example of such a star would be LP 261-75 \citep{irwin18}, which has an H$\alpha$ EW several times larger than that of TOI-2119. The shape of the H$\alpha$ profile is the result of self-absorption \citep{youngblood2022} and we do not see evidence that this is tied to the presence of the BD.

\subsection{Testing substellar isochrones}
\begin{figure}
    \centering
    \includegraphics[width=0.47\textwidth, trim={1.0cm 0.0cm 0.0cm 1.0cm}]{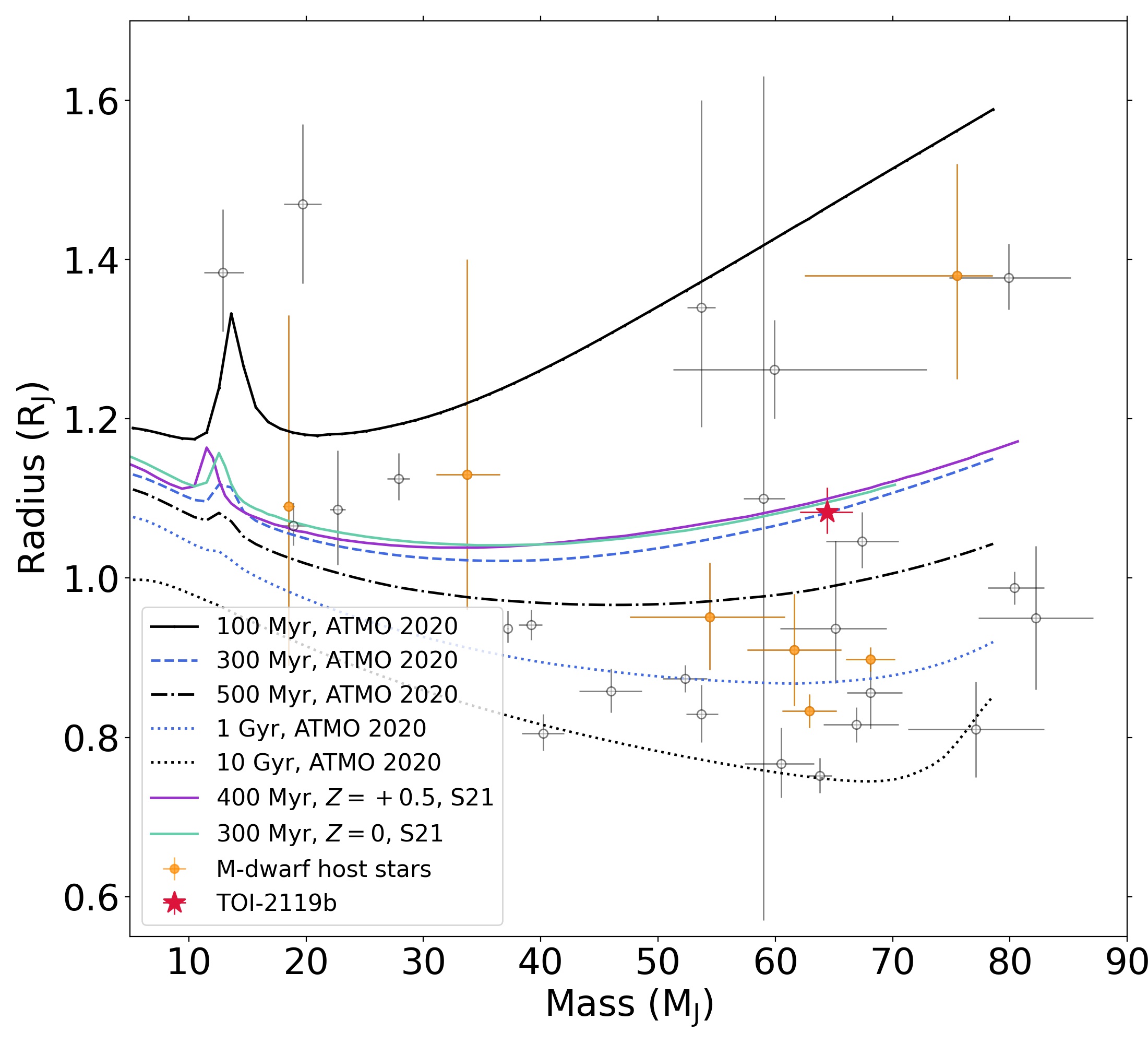}
    \caption{Mass--radius diagram for transiting brown dwarfs featuring substellar ATMO 2020 and S21 evolutionary models. Not pictured (at $R_b = 3.1{\rm R_J}$) is RIK 72b, which is a brown dwarf that transits an M-dwarf in Upper Scorpius (5-10 Myr old). There are a total of 9 transiting brown dwarf systems that involve an M-dwarf primary. Given our age estimate for the host star, we expect the TOI-2119 system to be $1.17 \pm 1.15$ Gyr old, which is consistent with the $300-400$ Myr solution for the companion BD, but not very precise given the relatively wide range of allowable ages. The best-fitting S21 models are shown in purple and light blue at different metallicities for the BD. Given the large uncertainty in the stellar age, this age range from $300-400$ Myr predicted by the S21 models is welcome.}
    \label{fig:mass-radius}
\end{figure}

Our determinations for the mass, radius, and age for TOI-2119b are $M_b = 64.4 \pm 2.3\mj$, $R_b= 1.08 \pm 0.03\rj$, and $1.17 \pm 1.15$ Gyr, respectively. We note here that a study by \cite{canas2021} independently verifies these mass and radius values. What we expect to see if the ATMO 2020 models accurately predict the evolution of a BD in a close orbit around an M-dwarf is that the models are consistent within roughly 1$\sigma$ of the mass and radius measurements of the BD at its age of 1.17 Gyr. What we see from Figure \ref{fig:mass-radius} is that this is the case, but primarily because of the large uncertainty with the age--rotation relationships used for TOI-2119.

\subsubsection{Sources of errors in the substellar radius determination}
Here we examine possible sources of error in our determination of the radius of the BD and host star as the radius is a critical part of our interpretation of the age--radius evolution of the BD as predicted by these models.

The ATMO 2020 and S21 models yield age estimates of 300 Myr at $Z=0$ or 400 Myr at $Z=+0.5$ for this BD (Figure \ref{fig:mass-radius}). These ages are certainly within the range of plausible ages for the host star at $1.17 \pm 1.15$ Gyr. If we assume the host star is instead a mid-type M-dwarf (M2.5--M6), then we find an age ($0.93\pm 0.24$ Gyr, Section \ref{subsec:ages}) that results in the BD and ATMO 2020 models being more consistent with each other. However, the mass of the M-dwarf is not consistent with a mid-type classification (generally $0.15 - 0.3\mst$).

We also consider the possibility of excess contribution in the $K_S$-band from the BD for our estimate of the host star's radius using Equation \ref{eq:rstar}. For the BD to have a radius that is consistent with the 1 Gyr ATMO 2020 evolutionary track (the track closest to the host star's age of 1.17 Gyr) in Figure \ref{fig:mass-radius}, the BD would have to be 10-15\% smaller depending on the metallicity ($0 \leq Z \leq 0.5$). This would mean that the BD is contributing an amount of light to the stellar SED such that the star appears larger than in reality (making the BD appear larger since its size is relative to that of the star). To correct for this hypothetical offset, the BD would need to contribute an additional $\Delta K = 0.5$ in Equation \ref{eq:rstar}. This means the BD would hypothetically have an apparent magnitude $K_S = 9.417$ that is blended with the star's light in this bandpass.

Based on the BD's mass ($M_b=64\mj$) and temperature ($T_{b, {\rm eff}}=2030$K), a plausible range of bolometric luminosities for this BD can be extrapolated from the mass--luminosity measurements in the BD sample explored in \cite{dupuy2017} and the S21 models. This range of luminosities is $-4.0 \leq \log{L_{\rm BD}/L_\odot \leq -3.5}$ and it can be approximated to a range of absolute $K_S$ magnitudes $10 \leq K \leq 12$ (following the same use of $K$ in Equation \ref{eq:mstar} and Equation \ref{eq:rstar}) using the sample presented in \cite{filippazzo2015}. This means that the BD can realistically contribute only an apparent magnitude $12.5 \leq K_S \leq 14.5$, which does not account for the hypothetical $\Delta K = 0.5$ in the stellar SED that would impact our measurement of the stellar radius.

Ultimately, the uncertainty on the host star's age (not its radius) is the limiting factor on how well TOI-2119b serves as a test point to the substellar mass--radius models we examine here.

\subsubsection{Temperature and atmosphere of the brown dwarf}
\begin{figure}
    \centering
    \includegraphics[width=0.47\textwidth, trim={0.5cm 0.0cm 0.0cm 0.0cm}]{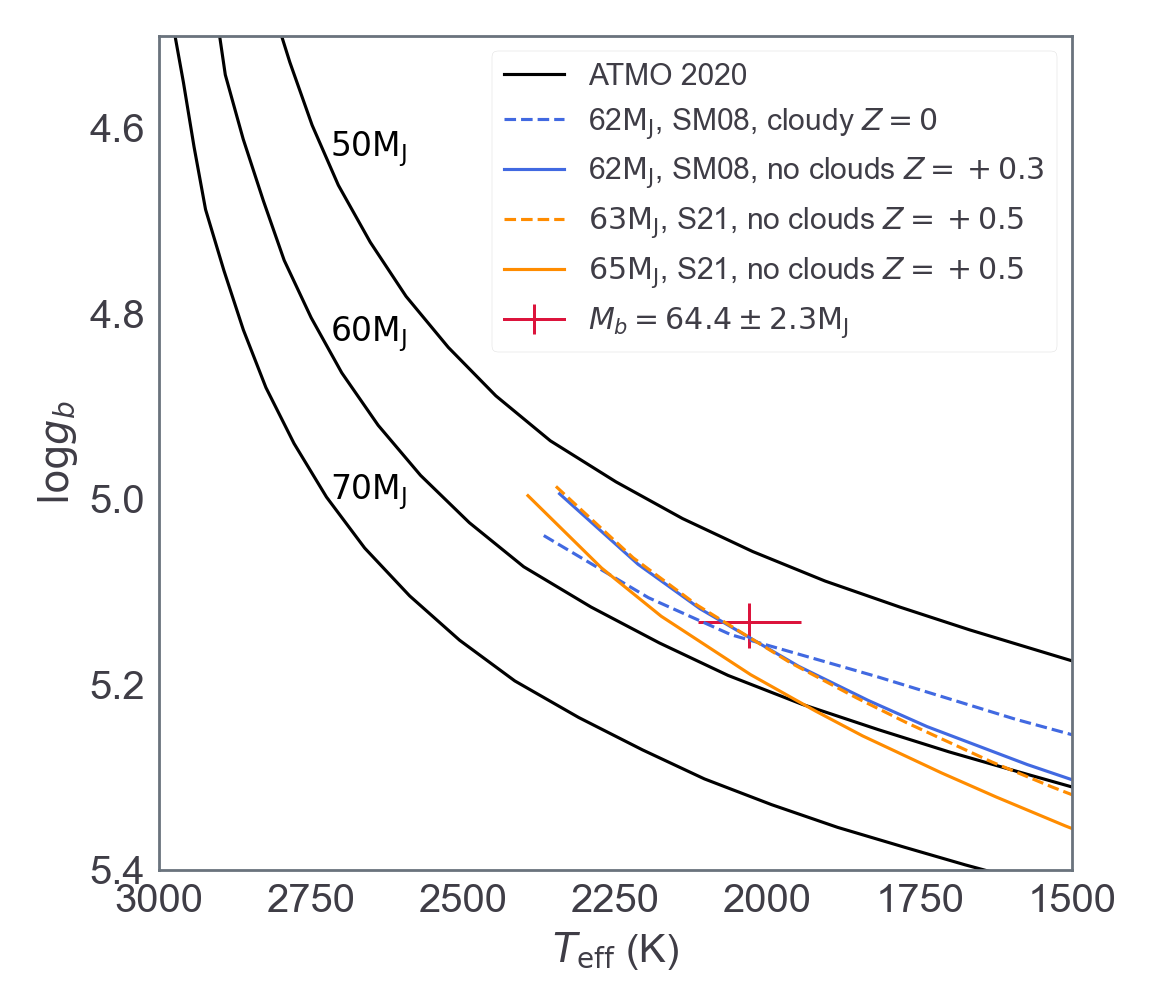}
    \caption{Substellar evolutionary tracks featuring the ATMO 2020, SM08, and S21 models. The ATMO 2020 models are only valid for $T_{b, {\rm eff}} < 2000$K because these models neglect some important sources of opacity at higher temperatures. The SM08 and S21 models are consistent with our $\log{g_b}$ and $T_{b, {\rm eff}}$ measurements of the BD, but given the intersection of these models at 2000K, it is not clear which (cloudy at solar metallicity vs. cloud-free and metal-rich) is favored. The ATMO 2020 models do not reproduce the data well, but this is expected given that the ATMO 2020 models are not valid at temperatures above 2000K.}
    \label{fig:bd_isochrone}
\end{figure}

The TOI-2119 system provides a rare opportunity to determine the temperature of the BD via the secondary eclipse events observed in the \tess\, data. This effective temperature for the BD is $T_{b, {\rm eff}}=2030 \pm 84$K and with it, we can examine the BD in a parameter space that utilizes mass, radius, and $T_{\rm eff}$ together. In Figure \ref{fig:bd_isochrone}, we compare the BD's $\log{g_b}$ and $T_{b, {\rm eff}}$ to the ATMO 2020, S21, and \cite{saumon08} (SM08) models. 

Based on the best-fitting models in Figure \ref{fig:bd_isochrone}, we believe that the BD is either metal-rich ($Z=+0.5$) with no clouds or cloudy with a relatively lower metallicity ($Z=0$). The latter scenario may be favored, if only slightly, if we assume that the BD formed out of the same disc material that the host formed out of and so, the two objects must have a very similar metallicity value ($\rm [Fe/H] = +0.06$ for the host star). \textcolor{black}{If instead we adhere more strictly to the best-fitting $\log{g_b}$--$T_{\rm eff}$ model from S21, then the cloud-free scenario is more favored. The significant overlap between the cloudy versus cloud-free S21 models with the uncertainties in the $\log{g_b}$, $T_{\rm eff}$, and mass of the brown dwarf exclude us from ruling either scenario out. Moreover, our inability to constrain the metallicity of the BD means that we cannot quantitatively invoke it here.}

\subsection{Mass ratio and formation scenarios}
As a consequence of their low mass, are M-dwarfs more prone to hosting BDs than more massive ($0.7-2.0\mst$) main sequence stars? How does the mass of the companion BD play a role in the formation and long-term stability of the BD? We cannot fully answer these questions here, but we can contextualise what we know about the population of M-dwarf--brown dwarf systems.

\subsubsection{M-dwarfs as host stars}
Approximately 27\% (9 of 33) of all known transiting BDs orbit an M-dwarf, with some systems involving an additional stellar companion \citep[e.g.][]{johnson11_bd, irwin10, jackman2019}. Though we remain in the small-number regime for the transiting BD population, this fraction of M-dwarf hosts to BDs is striking, especially when contrasted with the much smaller fraction ($\sim1-3\%$) of known transiting gas giant planets ($1\mj \leq M_b \leq 13\mj$) around M-dwarfs \citep[][and references therein]{hatzes15, mercer2020}. This discrepancy is likely a result of the different formation mechanisms at play between giant planets and BDs. It is expected that the majority of known transiting BDs formed via disc fragmentation or other similar star formation mechanisms given that disc fragmentation for the formation of a substellar companion is viable down to $4-10\mj$ \citep{schlaufman2018, tokovinin2020}. However, some BDs well above this mass range show evidence of formation via core accretion, like CWW 89Ab at $M_b=39\mj$ \textcolor{black}{showing evidence via a superstellar C/O ratio near unity} \citep{cww89a}, so the picture is not as clear when only considering the mass of the companion. This is why we must also consider other properties like the mass ratio $q = M_b/M_\star$ of the system.

It is not immediately obvious what range of mass ratios that is important to consider, but given that a quarter of the transiting BD population occupies a system with $q \geq 0.1$, we may start there. For reference, CWW 89Ab  serves as an example of a BD that formed via core accretion and it has a mass ratio of $q=0.03$ with its host star. So, the remaining work to be done is to, where possible, follow the methods in \cite{cww89a} by examining the reflected light or thermal emission from BDs in systems with $q \geq 0.1$ (usually only M-dwarfs or late K-dwarfs) in order to determine what link, if any, exists between the formation mechanisms and mass ratio in transiting BD systems.

\begin{figure}
    \centering
    \includegraphics[width=0.45\textwidth, trim={0.5cm 0.0cm 0.0cm 0.0cm}]{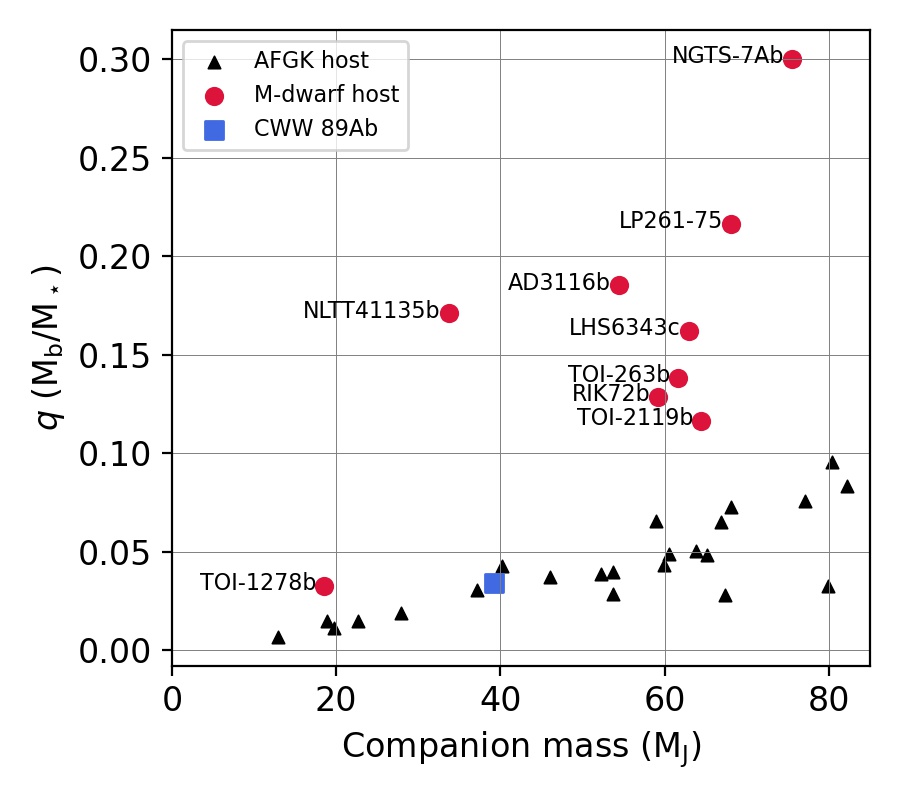}
    \caption{Mass ratio plotted against companion mass for transiting brown dwarf systems. \textcolor{black}{There is an apparent distinction in $q$ between brown dwarfs hosted by M-dwarfs and those hosted by non-M-dwarf stars.} \textcolor{black}{CWW 89Ab is shown for reference as a brown dwarf system with evidence of formation via core accretion and this system falls in the group of BDs with mass ratios $q \leq 0.1$.}}
    \label{fig:q}
\end{figure}

To work towards an understanding of this, we show how the transiting BD population behaves at different mass ratios versus \textcolor{black}{companion mass} in Figure \ref{fig:q}. With transiting BDs, we are limited to relatively short orbital periods given the nature of transit surveys and the decrease in transit probability with increasing orbital periods, \textcolor{black}{so any apparent trend seen now is subject to change with the addition of objects at wider separations. With the current population, there is a clear separation between those BDs hosted by M-dwarfs and those hosted by AFGK dwarf stars. From this trend we see in Figure \ref{fig:q}, one might infer that transiting brown dwarfs below a certain $q$ and mass form via core accretion while those above this threshold form via disc fragmentation or similar pathways. This may explain systems similar to TOI-1278b \citep[$q \approx 0.03$, $M_b=19\mj$,][]{artigau2021} that seem to follow the trend of non-M-dwarf host systems despite being hosted by an M-dwarf. The NLTT 41135 system \citep[$q \approx 0.17$, $M_b=34\mj$,][]{irwin10} is more difficult to interpret given that it is less massive than CWW 89Ab, which lends itself to an argument in favor of core accretion, but has a mass ratio six times larger (as well as one of the highest mass ratios among known transiting brown dwarfs). Again, given the relatively small number of BDs with M-dwarf hosts, we can do little more than qualitatively infer, but it is clear that a consideration of the mass ratio in addition to the BD mass is important in distinguishing the different formation scenarios at play in this population.} 

In the case of TOI-2119, we will rely on studies like \cite{tokovinin2020} to argue that this BD at a mass of $M_b=64\mj$, which is well above the range of efficient core accretion formation ($4-10\mj$) and above the mass of CWW 89Ab ($39\mj$), must have formed via disc fragmentation.

\subsubsection{Tidal evolutionary timescales}
We find a calculation of the circularization timescale $\tau_{\rm circ}$ for TOI-2119b to be largely uninformative of the formation and dynamical history of the system. TOI-2119b is in an eccentric orbit ($e=0.337$), so we know that it has not circularised, which is an observation consistent with the shortest possible circularization timescale $\tau_{\rm circ} \approx 15$ Gyr that we find using the relations in \cite{adams06}. Table \ref{tab:exofast_toi2119} quotes a $\tau_{\rm circ} = 154$ Gyr using a tidal quality factor $Q=10^6$, but values as low as $10^{4.5}-10^5$ are permitted for BDs \citep{cww89a}. Based on this range of $\tau_{\rm circ}$, the BD in this system may have formed in an eccentric, close-in orbit, or it may currently be undergoing a quick inward migration from a farther out formation distance that will take billions of years to complete. Both scenarios may be equally plausible given the large uncertainties in $\tau_{\rm circ}$ and the fact that the BD's orbit is eccentric. So, all that we can confirm is that the observed orbital eccentricity of TOI-2119b at its current age ($1.17\pm 1.15$ Gyr) is consistent with tidal evolutionary theory.

\subsection{Summary}
TOI-2119 is an early-type M-dwarf that hosts a transiting brown dwarf in an eccentric ($e=0.34$) orbital period of $P_{\rm orb} = 7.2$ days. The brown dwarf is completely eclipsed by the M-dwarf, allowing us to estimate the effective temperature of the brown dwarf to be $T_{b,{\rm eff}} = 2030$K. Using the scaling relations for M-dwarfs presented in \cite{mann2015} and \cite{engle2018}, we set priors on our global analysis for the star and BD, finding that the star has a mass $M_\star = 0.53 \pm 0.02\mst$, a radius $R_\star = 0.50 \pm 0.01\rst$, and an age (based on rotation-age relations for early-type M-dwarfs) of $1.17 \pm 1.15$ Gyr. The BD's mass and radius are $M_b = 64.4 \pm 2.3\mj$ and $R_b= 1.08 \pm 0.03\rj$, respectively. The age predicted by the substellar evolutionary models ($300-400$ Myr) for the brown dwarf is not inconsistent with the age estimate of the host star, though this stellar age estimate carries a large uncertainty with it. With improvements to the stellar age dating techniques, TOI-2119 may yet prove to be an important system in testing substellar evolutionary and atmospheric models at ages $<1$ Gyr. Until then, TOI-2119b serves as the newest brown dwarf amongst a rapidly growing population of substellar objects orbiting M-dwarf host stars.

\section*{Acknowledgements}
Funding for the \tess\, mission is provided by NASA's Science Mission directorate. This paper includes data collected by the \tess\, mission, which are publicly available from the Mikulski Archive for Space Telescopes (MAST). Resources supporting this work were provided by the NASA High-End Computing (HEC) Program through the NASA Advanced Supercomputing (NAS) Division at Ames Research Center for the production of the SPOC data products. We acknowledge the use of public TESS data from pipelines at the TESS Science Office and at the TESS Science Processing Operations Center.

This work makes use of observations from the LCOGT network. Part of the LCOGT telescope time was granted by NOIRLab through the Mid-Scale Innovations Program (MSIP). MSIP is funded by NSF.

This work is partly supported by JSPS KAKENHI Grant Number JP18H05439, JST CREST Grant Number JPMJCR1761, the Astrobiology Center of National Institutes of Natural Sciences (NINS) (Grant Number AB031010).

This work has made use of data from the European Space Agency (ESA) mission Gaia (\url{https://www.cosmos.esa.int/gaia}), processed by the Gaia Data Processing and Analysis Consortium (DPAC, \url{https://www.cosmos.esa.int/web/gaia/dpac/consortium}). Funding for the DPAC has been provided by national institutions, in particular the institutions participating in the Gaia Multilateral Agreement.

E. E-B. acknowledges financial support from the European Union and the State Agency of Investigation of the Spanish Ministry of Science and Innovation (MICINN) under the grant PRE2020-093107 of the Pre-Doc Program for the Training of Doctors (FPI-SO) through FSE funds.

TWC acknowledges the efforts of the members of the \tess\, Follow up Program and the Science Processing Operations Center in making the \tess\, data readily accessible for the analysis in this work.

For the purpose of open access, the author has applied a Creative Commons Attribution (CC BY) licence to any Author Accepted Manuscript version arising from this submission

\section*{Data Availability}
The light curve data underlying this article were accessed from the Mikulski Archive for Space Telescopes (MAST) at \url{https://mast.stsci.edu/portal/Mashup/Clients/Mast/Portal.html}. We strongly recommend the use of {\tt lightkurve} or similar software package for the management of any \tess\, data products. The derived light curve data generated in this research will be shared on reasonable request to the corresponding author. All data used in the analysis of the stellar and brown dwarf parameters are queried and properly formatted by {\tt EXOFASTv2} from their respective databases (see Section \ref{sec:exofast}). The installation and use instructions for {\tt EXOFASTv2} can be found at \url{https://github.com/jdeast/EXOFASTv2}.

\bibliographystyle{mnras}
\bibliography{citations}

\begin{table*}
\centering
\caption[]{MIST median values and 68\% confidence interval for TOI-2119, created using {\tt EXOFASTv2} commit number f8f3437. Here, $\mathcal{U}$[a,b] is the uniform prior bounded between $a$ and $b$, and $\mathcal{G}[a,b]$ is a Gaussian prior of mean $a$ and width $b$. The parameter $Age_{\rm EG18}$ indicates the age calculated from Equation \ref{eq:age}. Entries in the priors column labelled `Not modelled' are parameters that are not calculated by {\tt EXOFASTv2} specifically.} \label{tab:exofast_toi2119}

\begin{tabular}{lcccc}
~~~Parameter & Units & Priors & Values\\
\hline
\multicolumn{2}{l}{\bf Stellar Parameters:}&\smallskip\\
~~~~$M_*$\dotfill &Mass (\mst)\dotfill & $\mathcal{G}[0.522,0.009]$ &$0.525^{+0.020}_{-0.021}$\\
~~~~$R_*$\dotfill &Radius (\rst)\dotfill & $\mathcal{G}[0.499,0.014]$ &$0.500\pm0.015$\\
~~~~$L_*$\dotfill &Luminosity (\lst)\dotfill & - &$0.0397^{+0.0013}_{-0.0012}$\\
~~~~$\rho_*$\dotfill &Density (cgs)\dotfill & - &$5.97^{+0.36}_{-0.33}$\\
~~~~$\log{g}$\dotfill &Surface gravity (cgs)\dotfill & - &$4.763\pm0.018$\\
~~~~$T_{\rm eff}$\dotfill &Effective Temperature (K)\dotfill & $\mathcal{G}[3512,100]$ &$3621^{+48}_{-46}$\\
~~~~$[{\rm Fe/H}]$\dotfill &Metallicity (dex)\dotfill & $\mathcal{G}[-0.10,0.09]$ &$+0.055^{+0.084}_{-0.077}$\\
~~~~$Age$\dotfill &Age (Gyr)\dotfill & $\mathcal{G}[1.17,1.15]$ &$2.14^{+1.0}_{-0.90}$\\
~~~~$Age_{\rm EG18}$\dotfill &Age (Gyr)\dotfill & Not modelled &$1.17\pm 1.15$\\
~~~~$A_V$\dotfill &V-band extinction (mag)\dotfill & $\mathcal{U}[0,0.05116]$ &$0.076^{+0.064}_{-0.049}$ \\
~~~~$\sigma_{SED}$\dotfill &SED photometry error scaling \dotfill & - &$1.67^{+0.78}_{-0.45}$\\
~~~~$\varpi$\dotfill &Parallax (mas)\dotfill & $\mathcal{G}[31.765,0.026]$ &$31.765\pm0.026$\\
~~~~$d$\dotfill &Distance (pc)\dotfill & - & $31.481\pm0.026$\\
~~~~$P_{\rm rot}$\dotfill &Rotation period (days)\dotfill & Not modelled & $13.11 \pm 1.41$\\
~~~~$v\sin{I_\star}$\dotfill &Project equatorial velocity ($\rm km\,s^{-1}$)\dotfill & Not modelled &  $\leq 3.04 \pm 0.5$\\
\hline
\multicolumn{1}{l}{\bf Brown Dwarf Parameters:}& \\
~~~~$P_{\rm orb}$\dotfill & Period (days)\dotfill & - &$7.2008652\pm 0.0000018$\\
~~~~$M_b$\dotfill &Mass (\mj)\dotfill  & - &$64.4^{+2.3}_{-2.2}$\\
~~~~$R_b$\dotfill &Radius (\rj)\dotfill & - & $1.08\pm0.03$\\
~~~~$T_C$\dotfill &Time of conjunction (\bjdtdb)\dotfill & - &$2458958.678060\pm 0.00007$\\
~~~~$a$\dotfill &Semi-major axis (AU)\dotfill  & - &$0.0611\pm 0.0009$\\
~~~~$i$\dotfill &Orbital inclination (Degrees)\dotfill & - &$88.416^{+0.055}_{-0.057}$\\
~~~~$e$\dotfill &Eccentricity \dotfill  & - &$0.337^{+0.0019}_{-0.00064}$\\
~~~~$ecos{\omega_*}$\dotfill & \dotfill & - &$0.33641\pm 0.00045$\\
~~~~$esin{\omega_*}$\dotfill & \dotfill & - &$-0.014^{+0.019}_{-0.028}$\\
~~~~$\tau_{\rm circ}$\dotfill &Tidal circularization timescale (Gyr)\dotfill & \cite{adams06} &$152^{+21}_{-15}$\\
~~~~$\fave$\dotfill &Incident Flux (\fluxcgs)\dotfill & - &$0.01293^{+0.00052}_{-0.00049}$\\
~~~~$T_{eq}$\dotfill &Equilibrium temperature (K)\dotfill  & - &$502.4^{+5.0}_{-4.8}$\\
~~~~$K$\dotfill &RV semi-amplitude ($\rm m\, s^{-1}$)\dotfill  & - &$10270^{+230}_{-190}$\\
~~~~$R_b/R_*$\dotfill &Radius of planet in stellar radii \dotfill  & - &$0.22284^{+0.00065}_{-0.00067}$\\
~~~~$a/R_*$\dotfill &Semi-major axis in stellar radii \dotfill  & - &$26.26^{+0.71}_{-0.51}$\\
~~~~$\delta$\dotfill &Transit depth (fraction)\dotfill  & - &$0.04966\pm0.00030$\\
~~~~$\tau$\dotfill &Ingress/egress transit duration (days)\dotfill  & - &$0.02559^{+0.00045}_{-0.00043}$\\
~~~~$b$\dotfill &Transit Impact parameter \dotfill  & - &$0.6538\pm 0.067$\\
~~~~$logg_b$\dotfill &Surface gravity \dotfill & -  &$5.132^{+0.028}_{-0.020}$\\
~~~~$M_b\sin i$\dotfill &Minimum mass (\mj)\dotfill & -  &$64.4^{+2.3}_{-2.2}$\\
~~~~$M_b/M_*$\dotfill &Mass ratio \dotfill & -  &$0.1171^{+0.0034}_{-0.0030}$\\
\hline
\multicolumn{1}{l}{\bf Secondary eclipse Parameters:}& \\
~~~~$T_S$\dotfill &Time of eclipse (\bjdtdb)\dotfill & - &$2458956.5908^{+0.0018}_{-0.0017}$\\
~~~~$\delta_{S}$\dotfill &Measured eclipse depth (ppm)\dotfill  & - &$1053\pm88$\\
~~~~$\tau_S$\dotfill &Ingress/egress eclipse duration (days)\dotfill & - &$0.0242^{+0.0019}_{-0.0022}$\\
~~~~$b_S$\dotfill &Eclipse impact parameter \dotfill & - &$0.635^{+0.025}_{-0.033}$\\
~~~~$T_{\rm eff}$\dotfill &Brown dwarf effective temperature (K)\dotfill & Not modelled &$2030\pm84$\\
\hline
\multicolumn{2}{l}{Wavelength Parameters:}&&\tess\,\smallskip\\
~~~~$u_{1}$\dotfill &linear limb-darkening coeff \dotfill & -&$0.203\pm0.034$\\
~~~~$u_{2}$\dotfill &quadratic limb-darkening coeff \dotfill & -&$0.372\pm 0.036$\\
\\\multicolumn{2}{l}{Telescope Parameters:}&&TRES\smallskip\\
~~~~$\gamma_{\rm rel}$\dotfill &Relative RV Offset (m/s)\dotfill & -&$6940\pm110$\\
~~~~$\sigma_J$\dotfill &RV Jitter (m/s)\dotfill & -&$290^{+190}_{-110}$\\
\\\multicolumn{2}{l}{Transit Parameters:}&&TESS\smallskip\\
~~~~$\sigma^{2}$\dotfill &Added Variance \dotfill & -&$0.000001398\pm0.000000033$\\
~~~~$F_0$\dotfill &Baseline flux \dotfill & -&$1.000150\pm0.000028$\\
\hline
\end{tabular}
\end{table*}

\label{lastpage}

\end{document}